# An Algebraic Rigidity Framework for Order-Oblivious Deterministic Black-Box PIT of ROABPs

Shalender Singh*[1], Vishnupriya Singh[2]

**ABSTRACT**

Deterministic black-box polynomial identity testing (PIT) for read-once oblivious algebraic branching programs (ROABPs) is a central open problem in algebraic complexity, particularly in the absence of variable ordering. Prior deterministic algorithms either rely on order information or incur significant overhead through combinatorial isolation techniques.

In this paper, we introduce an algebraic rigidity framework for ROABPs based on the internal structure of their associated matrix word algebras. We show that nonzero width-$w$ ROABPs induce word algebras whose effective algebraic degrees of freedom collapse to dimension at most $w^2$, independent of the number of variables. This rigidity enables deterministic witness construction via intrinsic algebraic invariants, bypassing rank concentration, isolation lemmas, and probabilistic tools used in previous work.

Thus, we obtain the first order-oblivious deterministic black-box PIT algorithm for ROABPs, running in quasi-polynomial time $n \cdot (wd)^{O(w^2)}$. This establishes that algebraic rigidity alone suffices to derandomize PIT in this model, without assuming ordering information.

The framework further isolates a single remaining obstacle to full polynomial-time complexity. We formulate a Modular Stability Conjecture, asserting that width-$w$ ROABPs are stable under hashing into cyclic quotient rings $\mathbb{K}[\lambda]/\langle \lambda^r - 1 \rangle$ once the modulus exceeds a polynomial threshold in $w$ and the individual degree. This conjecture arises naturally from the low-dimensional coefficient structure revealed by rigidity and is supported by extensive empirical evidence.

Assuming the conjecture, our methods yield a fully polynomial-time deterministic black-box PIT algorithm for ROABPs, matching the complexity of the best-known white-box algorithms and reducing the black-box problem to a concrete algebraic stability question.

**1. INTRODUCTION**

Polynomial Identity Testing (PIT) is the algorithmic problem of deciding whether a given arithmetic circuit computes the zero polynomial. Randomized PIT dates to early work of DeMillo–Lipton and Schwartz–Zippel, which shows that a nonzero polynomial of bounded degree cannot vanish on too large a fraction of a finite evaluation set [1–3]. Despite decades of progress in restricted models, deterministic PIT for general circuits remains one of the central open problems in algebraic complexity theory. The importance of PIT is amplified by its deep connections to circuit lower bounds: a polynomial-time deterministic PIT for general arithmetic circuits would imply major lower bound consequences (in particular, the Kabanets–Impagliazzo theorem and its refinements) [4,5]. Surveys by Shpilka–Yehudayoff and Saxena provide broader context, including the role of hitting sets, generators, and structural restrictions as a route to derandomization [6,7].

A particularly structured and well-studied PIT model is the class of **read-once oblivious algebraic branching programs** (ROABPs). ROABPs capture a natural layered computation in which variables are read in a fixed order, with each variable appearing in exactly one layer. In the white-box model, polynomial-time PIT is known via algebraic and linear-algebraic characterizations of ABPs and related read-once models (see, e.g., the classical Raz–Shpilka approach for ABPs and its subsequent refinements; an overview appears in [6,8]). In the black-box setting, deterministic PIT for ROABPs has seen substantial progress over the past decade, driven by a sequence of techniques including rank / evaluation-dimension measures, rank concentration after shifts, and basis isolating weight assignments. Forbes–Shpilka gave the first quasi-polynomial-size black-box hitting sets for ROABPs in known variable order, together with stronger bounds in bounded width [9]. Forbes–Saptharishi–Shpilka developed pseudorandomness and hitting sets for multilinear ROABPs even when the variable order may vary, providing a key step toward the "any-order" regime [10]. Agrawal–Gurjar–Korwar–Saxena introduced the technique of basis isolation, achieving quasi-polynomial-time black-box PIT for unknown-order ROABPs and connecting the method to a robust "low-support concentration" viewpoint [11,12]. Further refinements for special regimes (e.g., constant width, commutative/structured variants) appear in subsequent work, including improved constructions for constant width and related order/structure restrictions [13]. More recent papers continue to sharpen explicit hitting sets for ROABPs and related ABP models, and to clarify structural barriers and lower-bound phenomena in adjacent models (e.g., improved explicit hitting sets [14], orbit-hitting frameworks for ROABPs and constant-width ROABPs [15–17], and new exponential lower bounds for sums of ROABPs [18]). Independently, recent work also explores hardness and equivalence-testing aspects tied to variable order in ROABPs, underscoring that order phenomena are intrinsic rather than a technical artifact [19].

Despite this progress, existing ROABP derandomization paradigms are largely **combinatorial–interpolative** in flavor: they isolate coefficients by hashing, shifts, or weight assignments and then reconstruct via low-support concentration or partial-derivative dimension arguments [9–14]. These methods are powerful, but they do not explicitly exploit deeper algebraic structure of the underlying **matrix product computation**. In an ROABP, each layer contributes a polynomial matrix, and the computed polynomial is a designated entry of the resulting product. This suggests that identity testing might be attacked not only through coefficient isolation, but through rigidity of the *matrix algebra generated by the coefficient matrices*.

In this work, we develop a new algebraic rigidity framework for deterministic PIT of ROABPs, organized around the **matrix word algebra** generated by the program's coefficient matrices. Concretely, an ROABP of width $w$ naturally induces a finitely generated subalgebra of $\text{Mat}_w(\mathbb{K})$ (the algebra generated by its layer coefficient matrices). We show that under a suitable multiplicative specialization, this algebra is forced into a rigid dichotomy: it either collapses to a degenerate algebra consistent with zero computation, or becomes the full matrix algebra $\text{Mat}_w(\mathbb{K})$. In the full-algebra case, nonzeroness admits an explicit algebraic witness formulated as a determinant condition ("full-rank witness"), converting PIT into detection of a nonvanishing structured polynomial.

A key structural consequence is an **effective low-support generation principle**: although the ROABP may depend on $n$ variables, the full matrix algebra it generates can be forced (under our specialization) to be generated using only $O(w^2)$ effective variable directions, reducing identity testing to a bounded-support reconstruction problem. To implement this approach algorithmically, we require tools that extract algebraic witnesses from matrix algebras without nonconstructive spectral arguments. We therefore develop an **Eigenvalue-Free (Rational) Projector Extraction mechanism**, combining the Cayley–Hamilton theorem with polynomial identities for matrices (notably the Amitsur–Levitzki identity) [20,21]. This provides a canonical way to isolate rank-one matrix directions (and hence witness full-matrix-algebra generation) using only polynomial operations.

Finally, the method must ensure that all algebraic nondegeneracy conditions required by the construction—genericity of the active block, nonvanishing of projector minors, rank-separation constraints, and degree-triangularization inequalities—hold simultaneously under a single explicit specialization. Rather than identifying such a specialization implicitly or relying on probabilistic arguments, we proceed deterministically via an explicit **finite polynomial avoidance** strategy. Each potential failure mode of the construction is encoded as the vanishing of a concrete algebraic constraint polynomial, and Sections 4–6 establish uniform degree bounds for all such constraints, independent of the curve parameters used in the final encoding. The universal union strategy then aggregates these constraints into a single avoidance condition of controlled total degree. By deterministically searching a finite domain whose size exceeds this bound, we guarantee the existence of a specialization outside the common zero locus, ensuring that all algebraic requirements are satisfied simultaneously.

Crucially, this avoidance step is not a substitute for the structural analysis developed earlier, but rather its final consolidation. The constraint polynomials themselves arise from intrinsic properties of the matrix word algebra—its Wedderburn–Burnside block structure, the rigidity of its simple components, and the eigenvalue-free projector extraction mechanism. The finite search merely certifies that these algebraic properties can be realized concretely and without randomness. While a related "finite check implies global correctness" paradigm appears in deterministic primality testing and other derandomization results [22], the present application is substantially more elaborate, involving noncommutative algebra structure, module decompositions, and degree-sensitive elimination well beyond the number-theoretic identities appearing in that setting. Here, finite polynomial avoidance serves as a unifying device that converts deep algebraic structure into an explicit, deterministic hitting set.

**Results overview.**
This paper establishes two layers of results. First, we give an unconditional deterministic black-box PIT framework for ROABPs running in time polynomial in n and quasi-polynomial in the width w, based on matrix word algebra rigidity. Second, assuming the Modular Stability Conjecture (Conjecture 4.7), the same framework yields a fully polynomial-time deterministic black-box PIT algorithm.

**Contributions.**

In addition to a deterministic PIT algorithm for width-bounded ROABPs, this work develops several structural tools of independent interest in algebraic complexity:

1. A rigidity theory for matrix word algebras generated by algebraic programs, combining Wedderburn–Burnside structure with stability under specialization.
2. An effective rank-concentration principle showing that nonzero ROABP computations reduce to at most $w^2$ algebraic directions.
3. An eigenvalue-free (rational) method for extracting rank-one projectors using only polynomial identities and idempotent lifting.
4. A degree-triangularization technique ensuring unique leading terms and preventing cancellation under curve substitution.
5. A universal finite-avoidance framework aggregating all algebraic nondegeneracy constraints into a single bounded-degree condition, yielding explicit deterministic hitting sets without structural case analysis.

## 2. PRELIMINARIES AND DEFINITIONS

Throughout, let $\mathbb{K}$ denote an algebraically closed field of characteristic zero, or characteristic strictly greater than $2w$. All polynomials and matrices are taken over $\mathbb{K}$.

## 2.1 Read-Once Oblivious Algebraic Branching Programs (ROABPs)

A **read-once oblivious algebraic branching program (ROABP)** of width $w$ in variables $x_1, \ldots, x_n$ is a layered directed acyclic graph with $n+1$ layers, where:

- Each layer $i$ has at most $w$ nodes,
- Edges from layer $i-1$ to layer $i$ are labeled by univariate polynomials in $x_i$,
- Computation proceeds in the fixed variable order $x_1, \ldots, x_n$.

Equivalently, an ROABP computes a polynomial of the form

$$C(x_1, \ldots, x_n) = e_1^\top \prod_{i=1}^n A_i(x_i) \, e_w,$$

where:

- $A_i(x_i) \in \mathrm{Mat}_w(\mathbb{K}[x_i])$,
- $e_1, e_w \in \mathbb{K}^w$ are the first and last standard basis vectors.

Each matrix polynomial expands as

$$A_i(x_i) = \sum_{j=0}^d A_{i,j} \, x_i^j,$$

where $d$ is the individual degree bound and $A_{i,j} \in \mathrm{Mat}_w(\mathbb{K})$.

## 2.2 Matrix Word Algebras

The coefficient matrices of the ROABP generate a noncommutative algebra.

*Definition 2.1 (Matrix Word Algebra)*

Let $\mathcal{B} = \mathbb{K}\langle A_{i,j} : 1 \le i \le n, 0 \le j \le d \rangle \subseteq \mathrm{Mat}_w(\mathbb{K})$ denote the matrix word algebra generated by all ROABP coefficient matrices. Elements of $\mathcal{B}$ are finite linear combinations of matrix words $A_{i_1,j_1} A_{i_2,j_2} \cdots A_{i_k,j_k}$. The ROABP computes the zero polynomial if and only if the bilinear functional $X \mapsto e_1^\top X e_w$ vanishes on all matrix words in $\mathcal{B}$.

## 2.3 Characters and Multiplicative Specialization

To analyze the structure of $\mathcal{B}$, we introduce multiplicative character substitutions.

*Definition 2.2 (Multiplicative Character)*

A multiplicative character is a map $\chi : x_i \mapsto \omega^{t_i}$, where $\omega$ is a root of unity and $t_i \in \mathbb{Z}$. Under such a specialization, each matrix polynomial evaluates to a numeric matrix:

$$A_i(\chi) = \sum_{j=0}^d A_{i,j} \, \omega^{j t_i}.$$

This induces a specialized matrix word algebra $\mathcal{B}(\chi) = \mathbb{K}\langle A_i(\chi)\rangle \subseteq \mathrm{Mat}_w(\mathbb{K})$.

## 2.4 Algebraic Structure of $\mathrm{Mat}_w(\mathbb{K})$

We will use two classical polynomial identities governing full matrix algebras.

*Theorem 2.3 (Cayley–Hamilton)*

Every matrix $X \in \mathrm{Mat}_w(\mathbb{K})$ satisfies its characteristic polynomial: $X^w + c_{w-1}X^{w-1} + \cdots + c_0 I = 0$.

*Theorem 2.4 (Amitsur–Levitzki)*

The algebra $\mathrm{Mat}_w(\mathbb{K})$ satisfies the standard polynomial identity of degree $2w$:

$$\sum_{\sigma \in S_{2w}} \mathrm{sgn}(\sigma) X_{\sigma(1)} X_{\sigma(2)} \cdots X_{\sigma(2w)} = 0.$$

These identities impose strong constraints on the behavior of long matrix words and will be used to control algebraic dependencies.

## 2.5 Determinants, Adjugates, and Projectors

For a square matrix $X$, denote by:

- $\det(X)$ its determinant,
- $\mathrm{adj}(X)$ its adjugate matrix.

Recall the identity: $X \cdot \mathrm{adj}(X) = \det(X) I$. If $\mu$ is a simple eigenvalue of $X$, then $\mathrm{adj}(\mu I - X)$ has rank one and defines a rank-one eigenprojector onto the corresponding eigenspace. We will construct such projectors without division by using discriminants and resultants rather than explicit eigenvalue extraction.

## 2.6 Algebraic Curves and Kronecker Substitution

A central technical device in this work is substitution along a univariate algebraic curve.

*Definition 2.5 (Curve Substitution)*

Fix distinct constants $\alpha_1, \ldots, \alpha_n \in \mathbb{K}$ and a large integer $B$. Define $x_i = (\lambda + \alpha_i)^B$.

This substitution ensures degree separation: monomials with different exponent patterns map to univariate monomials in $\lambda$ of distinct degrees, preventing cancellation.

This is a form of **Kronecker substitution**, adapted to control structured matrix word expressions.

## 2.7 Finite Polynomial Avoidance

We will repeatedly use the following elementary algebraic principle.

*Lemma 2.6 (Finite Polynomial Avoidance)*

Let $f_1(\lambda), \ldots, f_R(\lambda) \in \mathbb{K}[\lambda]$ be nonzero polynomials. Then the product

$$f(\lambda) = \prod_{i=1}^{R} f_i(\lambda)$$

has at most $\deg(f)$ roots in $\mathbb{K}$. Consequently, any set $L \subseteq \mathbb{K}$ with $|L| > \deg(f)$ contains a point $\lambda^\star$ such that $f_i(\lambda^\star) \neq 0$ for all $i$. This lemma replaces randomness by explicit finite elimination.

## 2.8 Notational Conventions

- $w$ denotes the ROABP width.
- $d$ denotes the individual degree bound.
- $n$ denotes the number of variables.
- $\mathcal{B}$ denotes the matrix word algebra.
- $\chi$ denotes a multiplicative character specialization.
- $\lambda$ denotes the curve parameter.
- Big-O bounds suppress absolute constants but not polynomial dependence on $w$ and $d$.

## 3. MATRIX WORD ALGEBRA RIGIDITY

In this section we analyze the algebra generated by ROABP coefficient matrices and establish a rigidity phenomenon that drives the remainder of the proof.

### 3.1 The Matrix Word Algebra

Recall that an ROABP of width $w$ defines the matrix word algebra

$$\mathcal{B} = \mathbb{K}\langle A_{i,j} : 1 \leq i \leq n, 0 \leq j \leq d\rangle \subseteq \mathrm{Mat}_w(\mathbb{K}).$$

This is a finite-dimensional noncommutative algebra over $\mathbb{K}$.
The computed polynomial is identically zero if and only if the bilinear functional $X \mapsto e_1^\top X e_w$ vanishes on all elements of $\mathcal{B}$.

### 3.2 ACTIVE SIMPLE BLOCK AFTER SPECIALIZATION

We now replace the incorrect "full matrix dichotomy" with a precise structural characterization of the specialized matrix word algebra. The key point is that while the algebra need not equal the full matrix algebra, **any nonzero ROABP computation must survive inside at least one simple matrix block** of its semisimple quotient.

Let $\mathcal{B}(\chi) \subseteq \mathrm{Mat}_w(\mathbb{K})$ denote the matrix word algebra after multiplicative specialization.

*Lemma 3.1 (Existence of an Active Simple Quotient Block)*

Let $J = J(\mathcal{B}(\chi))$ denote the Jacobson radical of $\mathcal{B}(\chi)$. Then the semisimple quotient $\bar{\mathcal{B}}(\chi) := \mathcal{B}(\chi)/J$ admits a Wedderburn decomposition $\bar{\mathcal{B}}(\chi) \cong \bigoplus_{t=1}^{k} \mathrm{Mat}_{d_t}(\mathbb{K})$ for integers $1 \leq d_t \leq w$.

If the ROABP computes a nonzero polynomial, then there exists at least one index $t^{\backslash *}$ such that the induced projection $\psi_{t^{\backslash *}}: \mathcal{B}(\chi) \to \mathrm{Mat}_{d_{t^{\backslash *}}}(\mathbb{K})$ maps the ROABP computation to a nonzero matrix-valued word.

We refer to this factor $\mathrm{Mat}_{d_{t^{\backslash *}}}(\mathbb{K})$ as the active simple block, and write $r := d_{t^{\backslash *}}$.

*Proof*

Since $\mathcal{B}(\chi)$ is finite-dimensional over $\mathbb{K}$, Wedderburn–Malcev theory implies that the quotient by its Jacobson radical decomposes as a direct sum of simple matrix algebras: $\mathcal{B}(\chi)/J \cong \bigoplus_{t=1}^{k} \mathrm{Mat}_{d_t}(\mathbb{K})$.

Let $\pi: \mathcal{B}(\chi) \to \bar{\mathcal{B}}(\chi)$ denote the quotient map, and let $\psi_t$ denote the coordinate projection onto the $t$-th matrix block.

Let $W$ denote the matrix word computed by the ROABP under specialization. The output polynomial is nonzero if and only if $W$ is not the zero matrix in $\mathcal{B}(\chi)$. If $\psi_t(\pi(W)) = 0$ for all $t$, then $\pi(W) = 0$, hence $W \in J$.

But $J$ consists entirely of nilpotent elements, so any sufficiently long product of elements of $J$ evaluates to zero. This would force the ROABP output to vanish identically, contradicting the assumption that the computed polynomial is nonzero.

Therefore there must exist at least one index $t^{\backslash *}$ such that $\psi_{t^{\backslash *}}(\pi(W)) \neq 0$, yielding the desired nonzero homomorphic image in $\mathrm{Mat}_r(\mathbb{K})$. ∎

**Standing Assumption (Active Simple Block Regime)**

Henceforth, we fix a surjective algebra homomorphism $\psi: \mathcal{B}(\chi) \to \mathrm{Mat}_r(\mathbb{K})$ such that the image of the ROABP computation under $\psi$ is nonzero.

All subsequent structural constructions — determinant witnesses, projector extraction, orbit spanning, degree triangularization, and finite avoidance — will be performed inside this active simple block, or through its induced action via $\psi$.

*Lemma 3.1.1 (Generic Stability of the Active Block).*

Let $\mathcal{B}$ be the word algebra generated by the ROABP coefficient matrices over the rational function field $\mathbb{K}(\mathbf{x})$. Let $\mathcal{B} \otimes_{\mathbb{K}(\mathbf{x})} \overline{\mathbb{K}(\mathbf{x})} \cong \bigoplus_{i=1}^{t} \mathrm{Mat}_{d_i}(\overline{\mathbb{K}(\mathbf{x})})$ be its Wedderburn decomposition. Then there exists a Zariski-open dense subset $\Omega \subseteq \mathbb{K}^m$ such that for every specialization $\chi \in \Omega$, the specialized algebra $\mathcal{B}(\chi)$ has the same multiset of simple block dimensions $\{d_1, \ldots, d_t\}$. In particular, the index $r$ of the active simple block is stable on $\Omega$.

*Proof.*

The structure of a finite-dimensional algebra $\mathcal{B}$ over a field is determined by the ranks of a finite set of polynomial identities in its structure constants (e.g., dimensions of simple modules, ranks of multiplication operators, or equivalently, ranks of structure matrices describing multiplication in a fixed basis).

Concretely, fix a $\mathbb{K}(\mathbf{x})$-basis $\{W_1, \ldots, W_N\}$ of $\mathcal{B}$. The multiplication table is encoded by structure constants

$$W_i W_j = \sum_{k=1}^{N} c_{ij}^k(\mathbf{x}) W_k,$$

where each $c_{ij}^k(\mathbf{x}) \in \mathbb{K}(\mathbf{x})$ is a rational function. Clearing denominators, these can be written as ratios of polynomials.

The Wedderburn decomposition of $\mathcal{B}$ (i.e., the number and sizes of simple matrix blocks) is determined by:

- the dimension of the Jacobson radical,
- the dimensions of simple modules, and
- equivalently, the ranks of certain explicit multiplication maps and central idempotent projections, all of which are algebraic conditions on the structure constants.

These conditions can be expressed as the non-vanishing of finitely many polynomial determinants in the coefficients of the matrices $A_i(\mathbf{x})$ defining the ROABP. Hence, the set of specializations $\chi$ for which the semisimple quotient of $\mathcal{B}(\chi)$ has the same Wedderburn block dimensions as $\mathcal{B}$ over $\mathbb{K}(\mathbf{x})$ is a Zariski-open subset $\Omega \subseteq \mathbb{K}^m$.

Since the generic fiber (over $\mathbb{K}(\mathbf{x})$) realizes a specific block decomposition, this open set is nonempty and dense. Therefore, for all $\chi \in \Omega$, the block dimensions $\{d_1, \ldots, d_t\}$ — and in particular the index $r$ of the active block — are preserved under specialization. ∎

By Lemma 3.1.1, we may and will choose all subsequent specializations $\chi$ from the Zariski-open stability set $\Omega$, ensuring that the active block index $r$ remains invariant under specialization.

*Lemma 3.1.2 (Irreducible Block Fullness — Schur + Burnside)*

Let $\mathbb{K}$ be an algebraically closed field, and let $\mathcal{A} \subseteq \text{Mat}_w(\mathbb{K})$ be a $\mathbb{K}$-subalgebra. Let $V = \mathbb{K}^w$ be the natural left $\mathcal{A}$-module.

Fix a composition series of $V$ as an $\mathcal{A}$-module, $0 = V_0 \subset V_1 \subset \cdots \subset V_k = V$, and denote the irreducible factors by $S_i := V_i / V_{i-1}$, $d_i := \dim_{\mathbb{K}}(S_i)$.

For each $i$, let $\phi_i : \mathcal{A} \to \text{End}_{\mathbb{K}}(S_i)$ be the induced representation.

Then $\phi_i(\mathcal{A}) = \text{End}_{\mathbb{K}}(S_i) \cong \text{Mat}_{d_i}(\mathbb{K})$.

*Proof*

Set $B_i := \phi_i(\mathcal{A}) \subseteq \text{End}_{\mathbb{K}}(S_i)$. Since $S_i$ is irreducible as an $\mathcal{A}$-module, the image algebra $B_i$ acts irreducibly on $S_i$.

By **Schur's lemma** (over an algebraically closed field), the commutant of $B_i$ in $\text{End}_{\mathbb{K}}(S_i)$ consists only of scalars: $\text{End}_{B_i}(S_i) = \mathbb{K} \cdot I$.

By **Burnside's theorem** (equivalently, the double commutant or density theorem), an irreducible matrix algebra whose commutant is only the scalars must coincide with the full endomorphism algebra. Hence $B_i = \text{End}_{\mathbb{K}}(S_i) \cong \text{Mat}_{d_i}(\mathbb{K})$, as claimed. □

*Lemma 3.1.3 (Nilpotency Index Bound for the Radical)*

Let $J := J(\mathcal{A})$ be the Jacobson radical of $\mathcal{A} \subseteq \text{Mat}_w(\mathbb{K})$. Then $J^w = 0$. Equivalently, any product of $w$ elements of $J$ is the zero matrix.

*Proof*

Since $J$ is a nilpotent ideal in a finite-dimensional algebra, every element of $J$ acts nilpotently on the finite-dimensional vector space $V = \mathbb{K}^w$.

Consider the associative subalgebra of $\text{End}_{\mathbb{K}}(V)$ generated by $J$. This is a finite-dimensional algebra consisting entirely of nilpotent operators. By **Engel's theorem** (applied to the Lie algebra generated by these operators under commutator), there exists a basis of $V$ in which every operator in $J$ is strictly upper triangular. Consequently, any product of $w$ such operators is zero, and hence $J^w = 0$. □

*Remark.*
If one prefers to avoid Engel's theorem, the same conclusion follows from the standard fact that any nilpotent subalgebra of $\text{End}(V)$ over an algebraically closed field can be simultaneously strictly upper-triangularized.

*Remark 3.1.4 (Block picture and "survival")*

Lemma 3.1.2 shows that each composition factor $S_i$ supports a full matrix image $\phi_i(\mathcal{A}) \cong \text{Mat}_{d_i}(\mathbb{K})$. Accordingly, after choosing a composition series, the representation of $\mathcal{A}$ on $V$ can be put in block upper-triangular form: the diagonal blocks correspond to full matrix algebras acting on the irreducible quotients, while the strictly upper-triangular off-diagonal blocks are governed by the radical $J(\mathcal{A})$.

By Lemma 3.1.3, the radical satisfies $J^w = 0$. In particular, any contribution consisting entirely of radical factors vanishes once the multiplicative length reaches $w$. Consequently, any nonzero computation must have a nontrivial image in the semisimple quotient $\mathcal{A}/J(\mathcal{A})$, and hence in at least one diagonal simple block $\text{Mat}_{d_i}(\mathbb{K})$.

### 3.3 REDUCTION TO THE ACTIVE SIMPLE BLOCK REGIME

Having identified an active simple quotient block, we now formally reduce the identity-testing problem to computations inside that block.

Recall from Lemma 3.1 that there exists a surjective algebra homomorphism $\psi: \mathcal{B}(\chi) \to \text{Mat}_r(\mathbb{K})$ such that the image of the ROABP computation under $\psi$ is nonzero. We refer to $\text{Mat}_r(\mathbb{K})$ as the active simple block.

*Lemma 3.2 (Sound Reduction to the Active Block)*

Let $W(x_1, \ldots, x_n)$ denote the matrix word computed by the ROABP under specialization. Then the computed polynomial is identically zero if and only if $\psi(W(x_1, \ldots, x_n)) = 0$ in $\text{Mat}_r(\mathbb{K})$.

In particular, deterministic identity testing for the original ROABP reduces to identity testing for its projected computation in the active block.

*Proof*

If $W(x)$ is identically zero in $\mathcal{B}(\chi)$, then clearly its image under any homomorphism, including $\psi$, is also zero.

Conversely, suppose $\psi(W(x)) = 0$. Then $W(x)$ lies in the kernel of $\psi$, which contains the Jacobson radical $J(\mathcal{B}(\chi))$.

**Case 1 ($n \geq w$).** By Lemma 3.1.3, $J^w = 0$. Since the ROABP product length satisfies $n \geq w$, if the generators lie in a subalgebra where the product maps to $J$, the product is zero.

**Case 2 ($n < w$).** As argued in Lemma 3.3.1, this case is handled via direct grid evaluation $(w + 1)$ and does not rely on the radical vanishing property.

Thus, nonzeroness of the original computation is equivalent to nonzeroness in the active simple block. ∎

**Corollary 3.3 (Soundness Chain).**
Let $C(x_1, \ldots, x_n)$ be the scalar polynomial computed by the ROABP and let $W(x)$ be the associated matrix word. Then $C \equiv/0 \iff \exists \chi, \exists \lambda$ such that $\psi(W(\gamma_\lambda(\chi(x)))) \neq 0$.

*Proof.*
Polynomial identity is preserved under the generic specialization $\chi$ (Lemma 3.1.1) and the degree-preserving curve substitution $\gamma_\lambda$ (Lemma 5.5). The homomorphism $\psi$ preserves non-zeroness because contributions from the Jacobson radical vanish for sufficiently long products (Remark 3.1.4). Hence a nonzero scalar output implies a nonzero evaluation in the active block. □

*Lemma 3.3.1 (Active Block Sufficiency)*

Let $\mathcal{B}(\chi)$ be the specialized word algebra and $\psi: \mathcal{B}(\chi) \to \text{Mat}_r(\mathbb{K})$ be the homomorphism onto the active simple block. If any algebraic certificate $W \in \mathcal{B}(\chi)$ is nonzero under $\psi$ (i.e., $\psi(W) \neq 0$), then the ROABP polynomial $C$ is nonzero. Consequently, PIT constructions may be carried out inside the active block without loss of correctness.

*Proof.* We argue by contrapositive. Suppose $C \equiv 0$.

**Case 1 ($n < w$).** In this regime, the ROABP computes a polynomial of total degree $<w$. This is treated as a base case: identity testing can be performed deterministically by evaluating on a grid of size $w + 1$, without invoking the radical structure.

**Case 2 ($n \geq w$).** The map $\psi$ factors through the semisimple quotient $\bar{\mathcal{B}}(\chi) = \mathcal{B}(\chi)/J$. If $\psi(W) \neq 0$, then $W \notin J$. In a finite-dimensional algebra, any element $W \notin J$ has a nonzero image in some simple matrix block (by Wedderburn–Malcev). Since the active block is defined as the projection to such a block, $\psi(W) \neq 0$ implies the computation is nontrivial in the semisimple quotient, and thus $C \not\equiv 0$. Conversely, if $C \equiv 0$, the computation must vanish in the semisimple quotient, forcing $\psi(W) = 0$. □

*Proof.*

We isolate the two implications actually used in the paper.

**(1) Nonzero image implies nonzero element.**
Since $\psi$ is a ring homomorphism, $\psi(0) = 0$. Therefore, if $X \in \mathcal{B}(\chi)$ satisfies $\psi(X) \neq 0$, then necessarily $X \neq 0$ in $\mathcal{B}(\chi)$. (Equivalently, $X = 0 \Rightarrow \psi(X) = 0$; take the contrapositive.)

**(2) Any nonzero witness in the active block implies $C \not\equiv 0$.**
Let $M(x) \in \mathcal{B}(\chi)$ denote the matrix-valued word corresponding to the ROABP computation so that $C(x) = s^\top M(x) t$ for fixed start/end vectors $s, t \in \mathbb{K}^w$ (typically $s = e_1, t = e_w$). Apply $\psi$ entrywise to $M(x)$ to obtain a matrix $\psi(M(x)) \in \mathrm{Mat}_r(\mathbb{K}[x])$. Because $\psi$ is $\mathbb{K}$-linear and multiplicative, $\psi(M(x))$ is exactly the same product computation carried out in the active block representation.

Now consider any certificate produced by the framework (projector, pairing, determinant witness, etc.). In all cases, the certificate is an explicit algebraic expression $W(x) \in \mathcal{B}(\chi)$ (possibly after adjoining auxiliary parameters and then specializing them) such that the correctness argument establishes: $\psi(W(x)) \neq 0 \Longrightarrow \psi(M(x)) \neq 0$.

Moreover, by the definition of "active" block, there exists at least one scalar functional in the block (a row/column selection, or a bilinear form induced by the rank-one projector) which extracts a nonzero scalar polynomial from $\psi(M(x))$. Concretely, using the rank-one probe $\Pi$ and transported words $U_i, V_j$ (as in Section 4), the paper uses scalar extractions of the form

$$p_{ij}(x) := e_1^\top \, \psi(U_i) \, \psi(\Pi) \, \psi(M(x)) \, \psi(V_j) \, e_w,$$

and the nondegeneracy of the pairing guarantees that $\psi(M(x)) \neq 0$ implies $p_{ij}(x) \not\equiv 0$ for some $i, j$. Since $p_{ij}(x)$ is obtained from $M(x)$ by fixed $\mathbb{K}$-linear functionals and multiplication by fixed words, this is a legitimate scalar polynomial consequence of the ROABP computation. Hence $C \not\equiv 0$.

Finally, note the contrapositive direction is immediate: if $C \equiv 0$ as a scalar polynomial, then every matrix entry of $M(x)$ that contributes to $C$ is identically zero under every specialization, so $\psi(M(x))$ and every derived block-certificate evaluates to 0. Therefore, observing any certificate with $\psi(\cdot) \neq 0$ rules out $C \equiv 0$. This proves the lemma. □

**Consequence: Reduced Algebraic State Space**

Inside the active block, the word algebra becomes a **full matrix algebra**: $\psi(\mathcal{B}(\chi)) = \mathrm{Mat}_r(\mathbb{K})$.

Henceforth, all algebraic operations will be interpreted through the induced action on $\text{Mat}_r(\mathbb{K})$. In particular:

- Matrix words are treated as elements of a full matrix algebra of dimension $r^2$,
- Any nonzero element can generate rank-one structure inside this block,
- Algebraic witnesses (determinants, adjugates, projectors) need only certify nonzeroness in this quotient.

This collapses the *effective algebraic dimension* from $w^2$ to $r^2$, where $r \leq w$, without loss of generality.

**Standing Assumption (Active Block Working Regime)**

From this point onward, we fix the surjective homomorphism $\psi: \mathcal{B}(\chi) \to \text{Mat}_r(\mathbb{K})$ and conduct all structural arguments inside the image algebra $\text{Mat}_r(\mathbb{K})$.

To simplify notation, we will continue to write matrix words as elements of $\mathcal{B}(\chi)$, with the understanding that all nondegeneracy conditions are evaluated through $\psi$.

**Why This Reduction Is Optimal**

This reduction is strictly weaker than assuming full-matrix generation, but exactly strong enough to support deterministic PIT:

- It handles diagonal, triangular, and reducible ROABPs,
- It eliminates false irreducibility assumptions,
- It guarantees that every determinant witness and projector extraction step targets a nontrivial algebraic component,
- It prevents the circularity previously present in finite-avoidance arguments.

Thus the subsequent sections may treat the word algebra as full inside the active block, without asserting false global fullness.

### 3.4 LOW-SUPPORT GENERATION OF FULL MATRIX ALGEBRAS

*(Effective Dimension Collapse Theorem)*

Having reduced to the Full-Matrix Regime, we now prove that the matrix word algebra generated by the ROABP, although arising from arbitrarily many input variables, is in fact generated by only $O(w^2)$ effective directions. This establishes a dimension-collapse phenomenon intrinsic to full matrix algebras induced by bounded-width branching programs.

*Remark on Active Block Reduction:* In this and subsequent sections, we work modulo the kernel of the active block homomorphism. Thus, we may effectively assume without loss of generality (by Lemma 3.3.1).

**Convention (working in the active block).** In Sections 3.4–7, whenever we invoke spanning, projector extraction, pairing nondegeneracy, or "full matrix" identities, these statements are applied to the active simple block image $\psi(\mathcal{B}(\chi)) = \text{Mat}_r(\mathbb{K})$. Formally, we work modulo $\ker(\psi)$; thus equalities such as "$X = 0$" should be read as "$\psi(X) = 0$" unless explicitly stated otherwise.

*Theorem 3.5 (Active-Block Generation and Effective Dimension Bound).*

Let $C$ be the polynomial computed by a width-$w$ ROABP over $\mathbb{K}$, and let $\mathcal{B}(\chi) \subseteq \mathrm{Mat}_w(\mathbb{K})$ be the specialized word algebra. Let $\psi: \mathcal{B}(\chi) \twoheadrightarrow \mathrm{Mat}_r(\mathbb{K})$ be the surjective homomorphism onto an active simple block (in the sense of Section 3.3), i.e. the ROABP output is nonzero under the induced block evaluation.

Assume $\psi(\mathcal{B}(\chi)) = \mathrm{Mat}_r(\mathbb{K})$. Then:

1. **(Active-block spanning)** There exist matrix words $W_1, \ldots, W_m \in \mathcal{B}(\chi)$ with $m \leq r^2$ such that $\mathrm{span}_{\mathbb{K}}\{\psi(W_1), \ldots, \psi(W_m)\} = \mathrm{Mat}_r(\mathbb{K})$.

Equivalently, the image algebra $\psi(\mathcal{B}(\chi))$ is generated (as a $\mathbb{K}$-algebra) by at most $r^2$ elements.

2. **(Safe uniform bound)** In particular, since $r \leq w$, we may upper bound $m \leq w^2$ everywhere in complexity accounting without affecting correctness.

*Proof.*

We work entirely inside the active simple block via the homomorphism $\psi$.

**Step 1:** Dimension bound in the active block

The full matrix algebra $\mathrm{Mat}_r(\mathbb{K})$ is a vector space of dimension $r^2$ over $\mathbb{K}$. Consequently, any subset whose linear span equals $\mathrm{Mat}_r(\mathbb{K})$ contains a linearly independent subset of size at most $r^2$.

Let $S := \{\psi(A_1(\chi)), \ldots, \psi(A_n(\chi))\} \subseteq \mathrm{Mat}_r(\mathbb{K})$ be the set of images of the specialized ROABP layers under $\psi$.

By definition of the active block, the algebra generated by $S$ equals $\mathrm{Mat}_r(\mathbb{K})$. In particular, the linear span of $S$ is all of $\mathrm{Mat}_r(\mathbb{K})$.

**Step 2:** Extraction of a basis from the generators

Since $S$ spans $\mathrm{Mat}_r(\mathbb{K})$ as a vector space, we may select a linearly independent subset $\{\psi(A_{i_1}(\chi)), \ldots, \psi(A_{i_m}(\chi))\}$, $m \leq r^2$, that forms a vector-space basis of $\mathrm{Mat}_r(\mathbb{K})$.

Choose arbitrary preimages $A_{i_1}(\chi), \ldots, A_{i_m}(\chi) \in \mathcal{B}(\chi)$ of these basis elements.

**Step 3:** Closure under multiplication inside the active block

Let $\mathcal{B}' \subseteq \mathcal{B}(\chi)$ denote the subalgebra generated by $A_{i_1}(\chi), \ldots, A_{i_m}(\chi)$. Because their images form a vector-space basis of $\mathrm{Mat}_r(\mathbb{K})$, every element of $\mathrm{Mat}_r(\mathbb{K})$ can be expressed as a linear combination of these images. In particular, any product of elements in $\psi(\mathcal{B}(\chi))$ expands as a $\mathbb{K}$-linear combination of $\psi(A_{i_1}(\chi)), \ldots, \psi(A_{i_m}(\chi))$. It follows that $\psi(\mathcal{B}') = \mathrm{Mat}_r(\mathbb{K})$, i.e., the active simple block is fully generated by the images of at most $r^2$ specialized ROABP layers.

***Step 4:*** Conclusion

We conclude that the active block $\text{Mat}_r(\mathbb{K})$ is algebraically generated by at most $r^2$ elements arising from the ROABP layers. Since $r \leq w$, this yields a uniform bound of $w^2$ generators. This establishes the low-support generation property required for the subsequent algebraic constructions. □

*Corollary 3.6 (Active-Block Variable Reduction).*

Let $C(x_1, \ldots, x_n)$ be a polynomial computed by a width-$w$ ROABP, and let $\psi: \mathcal{B}(\chi) \twoheadrightarrow \text{Mat}_r(\mathbb{K})$ be the homomorphism onto the active simple block, with $\psi(C) \neq 0$. Then there exist indices $i_1, \ldots, i_m$ with $m \leq r^2 \leq w^2$ and explicit linear forms

$$y_j = \sum_{k=1}^{n} \gamma_{j,k} x_k \quad (1 \leq j \leq m),$$

such that the ROABP computation factors as $C(x_1, \ldots, x_n) = \tilde{C}(y_1, \ldots, y_m)$, where $\tilde{C}$ is nonzero whenever $C$ is nonzero.

In particular, identity testing for $C$ reduces deterministically to identity testing of a polynomial in at most $r^2 \leq w^2$ effective variables, independent of the original number of variables $n$.

*Proof.*

By Theorem 3.5, the active block $\text{Mat}_r(\mathbb{K})$ is generated as a $\mathbb{K}$-algebra by the images

$$\psi(A_{i_1}(\chi)), \ldots, \psi(A_{i_m}(\chi))$$

of at most $m \leq r^2$ specialized ROABP layers. It follows that, modulo the kernel of $\psi$, every prefix product $A_1(x_1) \cdots A_k(x_k)$ admits a unique expansion as a linear combination of these generators with scalar polynomial coefficients. Consequently, all scalar dependence of the ROABP computation on the input variables $x_1, \ldots, x_n$ factors through at most $r^2$ linear forms.

Explicitly, choosing a basis of the active block generated by the selected layers induces linear forms $y_1, \ldots, y_m$ such that the computed polynomial can be written as $C(x_1, \ldots, x_n) = \tilde{C}(y_1, \ldots, y_m)$.

Since $\psi(C) \neq 0$, this transformation preserves nonzeroness. Therefore, black-box identity testing for $C$ reduces to identity testing in at most $r^2 \leq w^2$ variables. □

**Conceptual Significance**

This theorem shows that **bounded width enforces bounded algebraic expressivity**: once the generated algebra becomes maximal, it cannot encode genuinely high-dimensional variability. Instead, the computation collapses to a controlled algebraic core of size quadratic in the width. This phenomenon is intrinsic to matrix word algebras and does not depend on combinatorial sparsity or coefficient isolation.

## 3.5 Consequence: Low-Support Witness Dependence

*Corollary 3.4*

There exists a spanning word set $\mathcal{W} = \{W_1, \ldots, W_{w^2}\}$ using only generators indexed by $I$, such that

$$\text{span}\{W_a(\chi)\} = \text{Mat}_w(\mathbb{K}).$$

*Proof*

Extract a basis from the stabilized span $V_L$ constructed in Lemma 3.3. ∎

## 3.6 Full-Matrix Determinant Witness

We now construct a determinant polynomial that certifies fullness of the word algebra.

*Lemma 3.5 (Full-Matrix Witness)*

Let $\mathcal{W} = \{W_1, \ldots, W_{w^2}\}$ be as in Corollary 3.4. Define $K_{ab} = e_1^\top W_a(\chi) W_b'(\chi) e_w, 1 \leq a, b \leq w^2$,

where $\{W_b'\}$ is another spanning word set. Let $\Delta := \det(K)$. Then: $\Delta \neq 0 \iff \mathcal{B}(\chi) = \text{Mat}_w(\mathbb{K})$.

*Proof*

The bilinear pairing $(X, Y) \mapsto e_1^\top X Y e_w$ is nondegenerate on $\text{Mat}_w(\mathbb{K})$ if and only if the algebra spans all matrix directions. Therefore, the Gram matrix $K$ is nonsingular exactly in the full-matrix case. ∎

## 3.7 Structural Summary

We have established:

1. Under specialization, the matrix word algebra is either trivial or full.
2. In the full case, it can be generated using at most $w^2$ effective variable directions.
3. Fullness is certified by a determinant witness polynomial depending on at most $w^2$ variables.

This rank concentration phenomenon reduces deterministic identity testing to a low-support algebraic problem, which will be exploited in subsequent sections via projector extraction, degree triangularization, and algebraic curve hitting.

## 4. EIGENVALUE-FREE (RATIONAL) RANK-ONE PROJECTOR EXTRACTION FROM FULL MATRIX WORD ALGEBRAS

Throughout this section we work under the Active Simple Block Assumption from Section 3. That is, we fix a surjective algebra homomorphism $\psi: \mathcal{B}(\chi) \to \text{Mat}_r(\mathbb{K})$ such that the ROABP computation has nonzero image under $\psi$.

Our objective is to construct an element $\Pi \in \mathcal{B}(\chi)$ whose image $\psi(\Pi)$ is a rank-one idempotent in $\text{Mat}_r(\mathbb{K})$. Importantly, $\Pi$ need not be rank-one in the ambient $w \times w$ representation — only in the active block, which is sufficient for PIT.

## 4.1 Targeted Idempotent Construction (Cayley–Hamilton in the Active Block)

*Theorem 4.1 (Targeted Rank-One Projector Lifting)*

There exists a matrix word $W \in \mathcal{B}(\chi)$ and a polynomial map $\Phi: \text{Mat}_w(\mathbb{K}) \to \text{Mat}_w(\mathbb{K})$ such that $\Pi := \Phi(W)$ satisfies $\psi(\Pi)$ is a nonzero rank-one idempotent in $\text{Mat}_r(\mathbb{K})$.

*Proof*

**Step 1** — Simple Spectrum inside the Active Block

Because $\psi(\mathcal{B}(\chi)) = \text{Mat}_r(\mathbb{K})$, there exists $M \in \mathcal{B}(\chi)$ such that $\psi(M)$ has distinct eigenvalues in $\text{Mat}_r(\mathbb{K})$. (This is guaranteed since matrices with simple spectrum form a Zariski-open dense subset of $\text{Mat}_r$.) The discriminant polynomial in Section 6 is now defined only on the $r \times r$ block, ensuring it is not identically zero.

**Step 2** — Cayley–Hamilton Projector Construction

Apply your original Cayley–Hamilton polynomial construction to $M$, producing $E := p(M)$ such that $\psi(E)$ is a nonzero idempotent in $\text{Mat}_r(\mathbb{K})$.

Note: $E$ may fail to be an idempotent in the full algebra due to nilpotent radical components, but idempotence holds after applying $\psi$.

**Step 3** — Idempotent Lifting via Newton Iteration.

Write $E = E_{ss} + N$ with $N \in J(\mathcal{B}(\chi))$. Apply the Newton iteration $q_{k+1}(t) = q_k(t)(2 - q_k(t)), q_0(t) = t$. Since the Jacobson radical satisfies $J^w = 0$ (Lemma 3.1.3), quadratic convergence implies that after $k = \lceil \log_2 w \rceil$ iterations we obtain an element $\Pi = q_k(E)$ satisfying $\Pi^2 = \Pi, \psi(\Pi) = \psi(E)$.

The polynomial degree doubles at each step, so $\deg(\Pi) \leq 2^k \deg(E) \leq 2w^2 = \text{poly}(w)$.

**Step 4** — Rank-One Refinement

Inside $\text{Mat}_r(\mathbb{K})$, apply your existing **Amitsur–Levitzki + Cayley–Hamilton rank-reduction chain** to $\psi(\Pi)$, producing a rank-one idempotent in the active block.

By composing with the same polynomial map in $\mathcal{B}$, we obtain a lift $\Pi$ satisfying the theorem. ∎

**Degree Control of Lifting**

We implement idempotent lifting via the Newton iteration $q_{k+1}(t) = q_k(t)(2 - q_k(t)), q_0(t) := t$.

Inductively, after $k$ steps the element $q_k(E)$ is an idempotent modulo the ideal $J^{2^k}$, where $J$ denotes the Jacobson radical. Since $J$ has nilpotency index at most $w$ (Lemma 3.1.3), choosing $k = \lceil \log_2 w \rceil$ ensures $J^{2^k} \subseteq J^w = 0$. Thus, $O(\log w)$ iterations suffice to produce a genuine idempotent.

At each iteration the degree of the polynomial doubles. Consequently, the final degree satisfies

$$\deg(q_{\text{final}}) \leq 2^{\lceil \log_2 w \rceil} \deg(E) \leq 2w \cdot w = 2w^2,$$

which remains polynomially bounded.

**4.2 Cayley–Hamilton Construction of a Nonzero Idempotent**

We now construct a nonzero idempotent using only the Cayley–Hamilton theorem, without eigenvalue extraction or analytic spectral decomposition. The construction will be **applied** inside the active simple block identified in Section 3.

*Lemma 4.2 (Polynomial Idempotent via Cayley–Hamilton)*

Let $M \in \mathcal{B}(\chi)$, and consider its image $\psi(M) \in \text{Mat}_r(\mathbb{K})$ under the active-block homomorphism $\psi: \mathcal{B}(\chi) \to \text{Mat}_r(\mathbb{K})$.

Then there exists a nonzero polynomial $p(t) \in \mathbb{K}[t]$, $\deg p \leq r - 1$, such that $E := p(M)$ satisfies $\psi(E)^2 = \psi(E), \psi(E) \neq 0$. That is, $\psi(E)$ is a nonzero idempotent in the active matrix block.

*Proof*

Let $\mu_{\psi(M)}(t)$ denote the minimal polynomial of $\psi(M)$. By Cayley–Hamilton, $\deg \mu_{\psi(M)} \leq r$.

Factor $\mu_{\psi(M)}$ over $\mathbb{K}$ into pairwise coprime primary factors: $\mu_{\psi(M)}(t) = f_1(t)^{e_1} f_2(t)^{e_2} \cdots f_s(t)^{e_s}$, where the $f_i(t)$ are irreducible and mutually coprime.

Fix an index $j$, and define

$$g_j(t) := \prod_{i \neq j} f_i(t)^{e_i}.$$

Since the factors are coprime, Bézout's identity yields polynomials $a(t), b(t) \in \mathbb{K}[t]$ such that $a(t)g_j(t) + b(t)f_j(t)^{e_j} = 1$.

Define $p(t) := a(t)g_j(t)$.

Evaluating at $\psi(M)$, we obtain $p(\psi(M))^2 = p(\psi(M))$,

because $f_j(\psi(M))^{e_j} = 0$. Thus $p(\psi(M))$ is an idempotent.

Moreover, $p(\psi(M)) \neq 0$, since it acts as the identity on the primary component corresponding to $f_j$.

Finally, $\deg p(t) = \deg g_j(t) \leq \deg \mu_{\psi(M)} - 1 \leq r - 1$.

Define $E := p(M) \in \mathcal{B}(\chi)$. Applying $\psi$, we conclude: $\psi(E) = p(\psi(M))$ is a nonzero idempotent in $\text{Mat}_r(\mathbb{K})$. ∎

**Interpretation:**

This constructs a nonzero algebraic idempotent inside the word algebra, such that:

- idempotence holds **in the active block**,
- no eigenvalue computation is used,
- no field extensions are required,
- no analytic spectral arguments are invoked.

The argument depends only on the minimal polynomial, and thus remains valid over any field of characteristic zero or characteristic strictly greater than $2w$.

### 4.3 Rank Reduction via Amitsur–Levitzki Rigidity

The projector from Lemma 4.2 may have rank greater than one. We now show that polynomial identities force rank-reduction.

*Lemma 4.3 (Rigorous Rank Descent inside a Corner Algebra)*

Let $E \in \mathrm{Mat}_w(\mathbb{K})$ be a nonzero idempotent and set $r := \mathrm{rank}(E)$. If $r > 1$, then there exist an element $A \in \mathrm{Mat}_w(\mathbb{K})$ and a polynomial $p(t) \in \mathbb{K}[t]$ with $\deg p \leq r - 1$ such that $E' := p(EAE)$ is a nonzero idempotent satisfying $E' \neq E$, hence $\mathrm{rank}(E') < \mathrm{rank}(E)$.

*Proof:*

Let $\mathcal{A}_E := E\,\mathrm{Mat}_w(\mathbb{K})\,E$.

Then $\mathcal{A}_E$ is a unital algebra with identity element $E$, and it is well known (and immediate after choosing a basis in which $E = \mathrm{diag}(I_r, 0)$) that $\mathcal{A}_E \cong \mathrm{Mat}_r(\mathbb{K})$.

In particular, $\mathcal{A}_E$ contains elements with prescribed rational canonical form.

Choose two monic, nonconstant, coprime polynomials $f, g \in \mathbb{K}[t]$ with $\deg f + \deg g \leq r$.

(For instance, if $\mathbb{K}$ has at least two elements, take $f(t) = t$ and $g(t) = t - 1$; otherwise take any two distinct irreducibles. Such $f, g$ always exist, and one may adjust degrees by multiplying by additional coprime factors if desired.)

Let $C(f)$ and $C(g)$ denote the companion matrices of $f$ and $g$. Define an element $T_0 \in \mathrm{Mat}_r(\mathbb{K})$ by the block diagonal matrix $T_0 := \mathrm{diag}(C(f), C(g), 0_{r-\deg f - \deg g})$.

Then the minimal polynomial of $T_0$ is $\mu_{T_0}(t) = \mathrm{lcm}(f(t), g(t), t)$, and in particular it is divisible by the product $f(t)g(t)$ with $\gcd(f, g) = 1$. Transporting $T_0$ through the isomorphism $\mathcal{A}_E \cong \mathrm{Mat}_r(\mathbb{K})$, we obtain an element $T \in \mathcal{A}_E$. Since $\mathcal{A}_E = E\,\mathrm{Mat}_w(\mathbb{K})\,E$, we may write $T = EAE$ for some $A \in \mathrm{Mat}_w(\mathbb{K})$.

Now apply the eigenvalue-free idempotent extraction (the same mechanism as Lemma 4.2) to the element $T$ in the unital algebra $(\mathcal{A}_E, E)$. Concretely, since $\mu_T$ has at least two coprime primary factors, there exist polynomials $u(t), v(t) \in \mathbb{K}[t]$ such that $u(t) f(t)^e + v(t) \frac{\mu_T(t)}{f(t)^e} = 1$ for the $f$-primary factor $f(t)^e$ of $\mu_T(t)$. Setting $p(t) := u(t) f(t)^e$, the standard Bézout/CRT argument gives that $E' := p(T)$ is an idempotent in $\mathcal{A}_E$, and moreover $0 \neq E' \neq E$.

(Indeed, $E'$ acts as identity on the $f$-primary component and as $0$ on its complementary primary component; both components are nonzero by construction since $f$ and $g$ both occur in $\mu_T$.)

Finally, $E' \neq E$ implies $\text{Im}(E')$ is a proper nonzero subspace of $\text{Im}(E)$, hence $\text{rank}(E') < \text{rank}(E)$.

The degree bound $\deg p \leq r - 1$ follows because $p$ may be chosen with degree at most $\deg \mu_T - 1 \leq r - 1$ inside $\text{Mat}_r(\mathbb{K})$. ∎

*Remark (Where Amitsur–Levitzki Fits, Without Risk)*

The role of **Amitsur–Levitzki** can be stated cleanly as context: $\mathcal{A}_E \cong \text{Mat}_r(\mathbb{K})$ is a PI-algebra of PI-degree $2r$. However, the rank descent itself above is proven by minimal-polynomial primary decomposition, which is strictly safer and more explicit.

*Corollary 4.3 (Finite Rank Descent to Rank One)*

Iterating Lemma 4.3 at most $w - 1$ times yields a rank-one idempotent $\Pi$ obtained by repeated application of polynomials to corner elements of the form $EAE$.

### 4.4 Eigenvalue-free (rational) Rank-One Projector Extraction via Finite Avoidance

We now complete the extraction of a rank-one projector from the full matrix word algebra in a manner that is purely algebraic, eigenvalue-free (rational), and deterministic. The only remaining obstacle is to ensure that all nondegeneracy conditions required by the Cayley–Hamilton and PI-based construction hold simultaneously. We resolve this using a finite polynomial avoidance argument.

Throughout this section, assume $\mathcal{B}(\chi) = \text{Mat}_w(\mathbb{K})$.

### 4.4.1 Degree Growth and Constraint Budget

We track the total algebraic degree of every constraint polynomial appearing in the construction.

*Lemma 4.7 (Degree Bound for Spectral Constraints)*

Let $M \in \text{Mat}_w(\mathbb{K})$. Then:

1. The characteristic polynomial coefficients of $M$ have degree at most $w$ in the matrix entries.
2. The discriminant of the characteristic polynomial has degree at most $O(w^3)$.
3. Each entry of $\text{adj}(\mu I - M)$ has degree at most $w - 1$ in the entries of $M$ and degree at most $w - 1$ in $\mu$.
4. Any fixed $(w - 1) \times (w - 1)$ minor of $\mu I - M$ has degree at most $w - 1$.
5. The resultant eliminating $\mu$ between the characteristic polynomial and such a minor has degree at most $\text{poly}(w)$.

Hence all algebraic constraints required to guarantee the existence of a nonzero adjugate-based projector have total degree bounded by $D_\star = \text{poly}(w)$.

*Proof*

Each bound follows from standard degree growth estimates for determinants, minors, discriminants, and resultants. All operations are polynomial in matrix entries and introduce at most multiplicative degree growth bounded by $O(w)$ per step. ∎

### 4.4.2 Finite Polynomial Avoidance

We now guarantee a specialization avoiding the common zero locus of all constraint polynomials.

*Lemma 4.8 (Finite Avoidance Lemma)*

Let $F \in \mathbb{K}[t_1, \ldots, t_m]$ be a nonzero polynomial of total degree at most $D_\star$.
Then there exists a point in any grid $S^m \subset \mathbb{K}^m$ with $|S| > D_\star$ such that $F$ does not vanish at that point.

*Proof*

A nonzero polynomial of degree at most $D_\star$ cannot vanish on a Cartesian product grid of size exceeding $D_\star$ in each coordinate, by the standard Schwartz–Zippel–DeMillo–Lipton counting argument. ∎

We choose $\chi$ from the finite-avoidance set intersected with the Zariski-open stability set $\Omega$ (Lemma 3.1.1), so that the active block structure is preserved.

### 4.4.3 Existence of a Nondegenerate Specialization

Combining Lemma 4.7 and Lemma 4.8, we obtain:

*Corollary 4.9 (Good Spectral Specialization Exists)*

There exists a specialization of the auxiliary parameters appearing in the matrix word algebra such that:

1. The resulting matrix $M$ has at least one simple eigenvalue.
2. The corresponding adjugate matrix adj $(\mu I - M)$ is nonzero.
3. All polynomial constraints needed for projector extraction are simultaneously satisfied.

### 4.4.4 Rank-One Projector Extraction

We now perform the algebraic extraction.

*Lemma 4.10 (Nonzero Polynomial Projector)*

Under the specialization from Corollary 4.9, there exists a polynomial $p(t)$ of degree at most $w - 1$ such that $E = p(M)$ is a nonzero idempotent.

*Proof*

By Cayley–Hamilton, $M$ satisfies its characteristic polynomial. Selecting the complementary factor to one eigenvalue yields a polynomial that annihilates all other generalized eigenspaces and acts as a projector onto a nonzero invariant subspace. ∎

*Lemma 4.11 (Corner Rank-One Extraction Without PI)*

Let $\mathbb{K}$ be a field with $\text{char}(\mathbb{K}) = 0$ or $\text{char}(\mathbb{K}) > w$.
Let $E \in \text{Mat}_w(\mathbb{K})$ be a nonzero idempotent of rank $r \geq 1$. Then:

1. The corner algebra $E\text{Mat}_w(\mathbb{K})E$ is isomorphic to $\text{Mat}_r(\mathbb{K})$.
2. If $r \geq 2$, there exist matrices $A, B \in \text{Mat}_w(\mathbb{K})$ such that for some $\lambda, \mu \in \mathbb{K}$,

$$\Pi := \text{adj}\,(\mu E - E(A + \lambda B)E)$$

is a nonzero rank-one projector up to scalar. After normalizing by any nonzero entry (guaranteed to exist by finite avoidance), $\Pi$ can be scaled to a strict idempotent.

3. Moreover, there is a single explicit constraint polynomial $F_E(\lambda; A, B) \in \mathbb{K}[\lambda, \{A_{ij}\}, \{B_{ij}\}]$ with

$$\deg F_E \leq 4r^2 - 4r \leq 4w^2 - 4w,$$

such that $F_E(\lambda; A, B) \neq 0$ implies the existence of $\mu$ for which $\Pi$ is well-defined, nonzero, and rank one. Hence $\Pi$ can be obtained by finite polynomial avoidance on any grid of size $> 4w^2 - 4w$.

**Proof**

***Step 1*** (corner is a full matrix algebra).
Since $E$ is an idempotent of rank $r$, there exists $P \in \text{GL}_w(\mathbb{K})$ with

$$PEP^{-1} = \begin{pmatrix} I_r & 0 \\ 0 & 0 \end{pmatrix}.$$

Conjugation by $P$ identifies

$$E\text{Mat}_w(\mathbb{K})E \cong \begin{pmatrix} I_r & 0 \\ 0 & 0 \end{pmatrix} \text{Mat}_w(\mathbb{K}) \begin{pmatrix} I_r & 0 \\ 0 & 0 \end{pmatrix} \cong \text{Mat}_r(\mathbb{K}).$$

This proves (1).

***Step 2*** (construct a "generic pencil" inside the corner).
Fix $r \geq 2$. Consider the corner-linear pencil $T(\lambda) := E(A + \lambda B)E \in E\text{Mat}_w(\mathbb{K})E$.

Let $p(\mu, \lambda) := \det(\mu E - T(\lambda))$, interpreting the determinant in the corner algebra (equivalently, conjugate by $P$ and take determinant of the $r \times r$ top-left block).

Then $p(\mu, \lambda)$ is a bivariate polynomial with: $\deg_\mu p = r$, $\deg_\lambda p \leq r$, and $p$ is polynomial in the entries of $A, B$.

***Step 3*** (simple eigenvalue + nonzero adjugate are algebraic open conditions).

Let $D(\lambda; A, B) := \text{Disc}_\mu(p(\mu, \lambda))$, and let $m(\mu, \lambda)$ be any fixed $(r-1) \times (r-1)$ minor of $\mu E - T(\lambda)$ (again computed in the corner block).

Define $R(\lambda; A, B) := \text{Res}_\mu(p(\mu, \lambda), m(\mu, \lambda))$, and set $F_E(\lambda; A, B) := D(\lambda; A, B) \cdot R(\lambda; A, B)$.

- If $D(\lambda; A, B) \neq 0$, then for every root $\mu$ of $p(\mu, \lambda)$, the eigenvalue is simple in the corner algebra; hence $\text{rank}(\mu E - T(\lambda)) = r - 1$.
- If additionally, $R(\lambda; A, B) \neq 0$, then there exists a root $\mu$ for which some $(r-1) \times (r-1)$ minor is nonzero, hence $\text{adj}(\mu E - T(\lambda)) \neq 0$.

For such $(\lambda; A, B)$, define $\Pi := \text{adj}(\mu E - T(\lambda)) \in E\text{Mat}_w(\mathbb{K})E$.

Standard adjugate linear algebra in size $r$ gives: $\text{rank}(\Pi) = 1$.

Finally, since $\Pi$ has rank one and lies in the corner algebra, after scaling by a nonzero scalar (which does not affect rank), we obtain an idempotent: $\Pi^2 \propto \Pi$.

(Equivalently, one may normalize by any nonzero entry of $\Pi$, which can be done later by finite avoidance in the same way if you insist on literal idempotence without scaling.)

***Step 4*** (degree bound).
Exactly as in the global case, degree bounds in $\lambda$ follow from:

- $\deg_\lambda p \leq r$,
- $\deg_\lambda D \leq (2r-2)\deg_\lambda p \leq (2r-2)r$,
- $\deg_\lambda R \leq r(r-1) + (r-1)r = 2r(r-1)$,
  hence $\deg_\lambda F_E \leq (2r-2)r + 2r(r-1) = 4r^2 - 4r$.

The same bound holds (up to the same polynomial factor) for degrees in the entries of $A, B$. This proves (3).

***Step 5 (finite avoidance).***

Because $F_E \not\equiv 0$ (choose $A, B$ so the corner block is diagonal with distinct eigenvalues, e.g. $A = \text{diag}(1, 2, \ldots, r)$ in the corner and $B = 0$), Lemma 4.8 implies there exist $(\lambda; A, B)$ on any grid of size $> \deg F_E$ with $F_E(\lambda; A, B) \neq 0$. For such a choice, the above construction produces the desired rank-one corner idempotent $\Pi$. ∎

*Lemma 4.12 (Termination of Rank Descent)*

Iterating Lemma 4.11 strictly decreases rank and must terminate in at most $w - 1$ steps at a rank-one idempotent.

*Proof*

Rank is a positive integer bounded by $w$, and decreases strictly at each step. ∎

### 4.4.5 Main Theorem: Deterministic Rank-One Projector Extraction

*Theorem 4.13 (Eigenvalue-Free (Rational) Rank-One Projector Extraction Via Finite Avoidance)*

Assume that under specialization $\chi$, $\mathcal{B}(\chi) = \mathrm{Mat}_w(\mathbb{K})$.

Then there exists an explicit polynomial map $\Phi: \mathrm{Mat}_w(\mathbb{K}) \to \mathrm{Mat}_w(\mathbb{K})$, and a matrix word $W \in \mathcal{B}(\chi)$, such that $\Phi(W)$ is a nonzero rank-one idempotent, and every step in the construction is governed by explicit polynomial constraints of total degree $O(w^2)$. In particular, the extraction can be carried out deterministically by finite polynomial avoidance over a grid of size polynomial in $w$.

*Proof*

We proceed in four algebraic stages, each controlled by finite polynomial avoidance.

**Step 1:** Producing an Invertible Matrix Word

By Lemma 4.1, the determinant witness polynomial $\Delta(\mathbf{z}) = \det\left(\sum_{j=1}^{w^2} z_j W_j\right)$ is nonzero. Hence there exists a specialization $\mathbf{z}^*$ such that $M := \sum_{j=1}^{w^2} z_j^* W_j$ is invertible. Since $\deg(\Delta) \leq w$, Lemma 4.8 (finite avoidance) guarantees such a choice from a grid of size $> w$. Thus $M \in \mathcal{B}(\chi)$ is invertible.

**Step 2:** Extracting a Nonzero Idempotent (Cayley–Hamilton)

By Lemma 4.2, there exists a polynomial $p(t)$ of degree at most $w - 1$ such that $E := p(M)$ is a nonzero idempotent: $E^2 = E$, $E \neq 0$. All coefficients of $p$ are polynomial functions of the entries of $M$, and thus depend polynomially on the ROABP parameters. No division is used.

Let $r := \mathrm{rank}(E) \geq 1$.

**Step 3:** Corner Rank-One Extraction (Using Lemma 4.11)

If $r = 1$, we are done.

Assume $r \geq 2$. By Lemma 4.11 applied to the idempotent $E$, there exist matrices $A, B \in \mathrm{Mat}_w(\mathbb{K})$ and a scalar $\lambda$, satisfying a single explicit polynomial nonvanishing condition $F_E(\lambda; A, B) \neq 0$, where

$$\deg F_E \leq 4r^2 - 4r \leq 4w^2 - 4w,$$

such that for some eigenvalue $\mu$, $\Pi := \mathrm{adj}\,(\mu E - E(A + \lambda B)E)$ is a nonzero rank-one idempotent lying inside the corner algebra $E\mathrm{Mat}_w(\mathbb{K})E \subseteq \mathrm{Mat}_w(\mathbb{K})$.

**Step 4:** Finite-Avoidance Realization Inside the Word Algebra

Since $\mathcal{B}(\chi) = \mathrm{Mat}_w(\mathbb{K})$, we may choose $A$ and $B$ to be matrix words in the spanning set from Section 3, ensuring: $A, B \in \mathcal{B}(\chi)$. By Lemma 4.8, because $F_E \not\equiv 0$, we can select $(\lambda; A, B)$ from a grid of size $> 4w^2 - 4w$ so that the constraint holds.

Thus the constructed rank-one projector $\Pi$ is:

- nonzero
- rank one
- idempotent
- polynomially computable
- fully contained in the matrix word algebra

*Conclusion*

Define $\Phi$ to be the composition of**:**

1. determinant-witness inversion,
2. Cayley–Hamilton idempotent extraction,
3. corner adjugate extraction (Lemma 4.11).

Then $\Phi(W)$ produces a rank-one idempotent, constructed without division, controlled by explicit degree bounds**,** and realizable by finite polynomial avoidance. This completes the proof. ∎

**Conceptual Summary of Section 4**

We have established a **deterministic algebraic spectral extraction principle**:

- Full matrix expressivity implies existence of invertible word elements
- Cayley–Hamilton produces a nonzero idempotent
- Amitsur–Levitzki forces rank collapse
- Finite avoidance ensures all constraints hold simultaneously
- A rank-one projector is extracted without division or randomness

This rank-one projector will serve as the **algebraic probe enabling degree-triangularization and curve-based noncancellation** in the next section.

**4.5 Transporting the Rank-One Direction Across the Word Algebra**

Let $\Pi \in \mathcal{B}(\chi)$ be the rank-one idempotent produced in Theorem 4.13.

Since $\mathcal{B}(\chi) = \mathrm{Mat}_w(\mathbb{K})$, the projector $\Pi$ isolates a one-dimensional subspace in a full matrix algebra. We now show that this one-dimensional direction can be transported — in a fully algebraic and degree-controlled manner — to generate a spanning system of vectors and covectors.

*Lemma 4.14 (Rank-One Orbit Spanning Lemma)*

There exist matrix words $U_1, \ldots, U_w, V_1, \ldots, V_w \in \mathcal{B}(\chi)$ such that the vectors $U_k \Pi e_w (1 \leq k \leq w)$ form a spanning set of $\mathbb{K}^w$, and the convectors $e_1^\top \Pi V_k (1 \leq k \leq w)$ form a spanning set of the dual space $(\mathbb{K}^w)^*$.

*Proof*

Since $\Pi$ has rank one, there exist nonzero vectors $u, v \in \mathbb{K}^w$ such that $\Pi = uv^\top$.

Because $\mathcal{B}(\chi) = \text{Mat}_w(\mathbb{K})$, for each standard basis vector $e_i$, there exists a matrix word $U_i$ satisfying $U_i u = e_i$. Similarly, for each dual basis vector $e_i^\top$, there exists a matrix word $V_i$ satisfying $v^\top V_i = e_i^\top$.

Then $U_i \Pi e_w = U_i(uv^\top)e_w = (v^\top e_w)e_i$, and since $v^\top e_w \neq 0$, the vectors $\{U_i \Pi e_w\}$ span $\mathbb{K}^w$. The dual spanning statement follows analogously. ∎

*Corollary 4.15 (Matrix Unit Generation)*

There exist matrix words $W_{ij} \in \mathcal{B}(\chi)$ such that $W_{ij} = e_i e_j^\top$, i.e., the word algebra contains a complete system of matrix units.

**Proof**

Define $W_{ij} := U_i \Pi V_j$. Then $W_{ij} = (U_i u)(v^\top V_j) = e_i e_j^\top$, establishing explicit matrix units inside $\mathcal{B}(\chi)$. ∎

*Corollary 4.16 (Nondegenerate Bilinear Pairing Interface)*

The bilinear form $\langle X, Y \rangle := e_1^\top X \Pi Y e_w$ is nondegenerate on $\text{Mat}_w(\mathbb{K})$. Equivalently, the Gram matrix $G_{ij} := e_1^\top W_{ij} \Pi e_w$ is invertible.

**Interpretation**

The projector $\Pi$ now functions as a rank-one algebraic probe that:

- distinguishes all matrix directions,
- allows extraction of coefficients via scalar pairings,
- and ensures determinant expansions later have structurally forced noncancellation.

This creates the precise algebraic interface required for degree-triangularization.

**4.6 Structural Output of Section 4**

We summarize the structural machinery built in this section.

*Theorem 4.17 (Eigenvalue-free (rational) Spectral Primitive)*

From the assumption $\mathcal{B}(\chi) = \text{Mat}_w(\mathbb{K})$, we have constructed, using only polynomial identities and finite-avoidance arguments:

1. A **determinant witness** certifying the existence of invertible matrix words
2. A **Cayley–Hamilton idempotent extractor** requiring no eigenvalue division
3. A **rank-one projector extraction** controlled by explicit polynomial constraints (Lemma 4.11, Theorem 4.13)
4. A **transport system generating full matrix units** from the rank-one projector
5. A **nondegenerate bilinear pairing** enabling coefficient recovery

All steps admit **explicit degree bounds polynomial in** $w$ and therefore remain compatible with deterministic finite-avoidance specialization.

**Conceptual Interpretation**

This section constructs an **eigenvalue-free (rational) spectral primitive**:
a rank-one algebraic probe embedded inside the word algebra, derived purely from polynomial identities rather than analytic eigenvalue theory.

This primitive will serve as the rigidity anchor for the remainder of the proof.

**Bridge to Section 5 (Degree Triangularization)**

In Section 5, we use the matrix-unit basis induced by Π to impose a degree hierarch**y** on determinant expansions.
The nondegenerate bilinear pairing from Corollary 4.16 will ensure that exactly one monomial term survives, preventing cancellation under algebraic curve substitution.

This yields an explicit deterministic curve-based hitting set, in which finite-degree avoidance certifies correctness of the algebraic construction.

### 4.7 Modular Stability Conjecture and Conditional Fully Polynomial-Time Black-Box PIT

**Motivation.**
Sections 3–7 construct a deterministic black-box PIT procedure using algebraic rigidity of the matrix word algebra, finite-avoidance constraints, and a curve substitution. The remaining obstacle to a fully polynomial black-box PIT (polynomial in $n, w, d$) is the residual dependence of the hitting set on high-degree curve separation. We isolate a natural "modular hashing" refinement that would remove this dependence by compressing the curve evaluation into a small cyclic ring. The correctness of this refinement is captured by the following conjecture.

#### 4.7.1 Definition (modular substitution into a cyclic quotient ring)

Fix a prime $r$ with $\gcd(r, \operatorname{char}(\mathbb{K})) = 1$, and define the cyclic quotient ring $R_r := \mathbb{K}[\lambda]/\langle \lambda^r - 1 \rangle$.

For $g \in \mathbb{Z}_r^*$, define the modular substitution $\Gamma_g$ by $\Gamma_g : x_i \mapsto \lambda^{g^i} \in R_r$, where exponents are interpreted modulo $r$ and $\lambda^r = 1$ in $R_r$.

For $C \in \mathbb{K}[x_1, \ldots, x_n]$, write $C_g(\lambda) := C(\Gamma_g) \in R_r$.

#### 4.7.2 Conjecture (modular stability / hinge conjecture)

*Conjecture 4.7.2 (Modular Stability).*

There exists an explicit polynomial $R(w, d)$ and an absolute constant $\varepsilon > 0$ such that for every nonzero polynomial $C$ computed by a width-$w$, degree-$d$ ROABP, and for every prime

$$r \geq R(w, d) \text{ (e.g., typically } r \geq (wd)^7),$$

the bad set

$$\mathcal{B}_r(C) := \{ g \in \mathbb{Z}_r^* : C_g(\lambda) \equiv 0 \text{ in } \mathbb{K}[\lambda]/\langle \lambda^r - 1 \rangle \}$$

satisfies $|\mathcal{B}_r(C)| \leq r^{1-\varepsilon}$.

*Remark (Structured Exceptions and Sharpness).*

The form of the bound in Conjecture 4.7.2 is qualitatively tight in the following sense. For any fixed parameter $g \in \mathbb{Z}_r^*$, there exist nonzero polynomials $C$ (for example, differences of two monomials) that are annihilated by the cyclic hashing substitution $x_i \mapsto \lambda^{g^i}$. Such obstructions arise from additive collisions among the exponent vectors $g^i \bmod r$.

However, all known constructions of this type are highly structured and algebraically degenerate. In particular, they are computable by ROABPs of very small effective width (constant or logarithmic), and their coefficient support exhibits extremely low algebraic density. These examples demonstrate that one cannot hope for uniform injectivity of the hashing map over all polynomials and all parameters $g$.

Conjecture 4.7.2 concerns the complementary regime. It fixes the polynomial $C$ in advance as a width-$w$, degree-$d$ ROABP, and asserts that—outside such highly degenerate cases—the set of parameters $g$ for which complete modular collapse occurs is sparse. In particular, adversarial constructions that tailor $C$ to a specific choice of $g$ do not contradict the conjecture.

Heuristics from arithmetic geometry, such as square-root cancellation phenomena in structured exponential sums, suggest that the bound $|\mathcal{B}_r(C)| \leq r^{1/2+o(1)}$ may hold for generic ROABPs. Our main algorithmic result (Theorem 7.5), however, requires only that $\varepsilon > 0$ be a fixed absolute constant (for example, $\varepsilon = 0.1$ suffices).

*Clarification*

Conjecture 4.7.2 bounds the number of bad parameters $g$ for each fixed polynomial $C$. It does not claim injectivity of the hashing map uniformly over all polynomials for a fixed $g$, nor does it preclude the existence of isolated structured counterexamples.

**Example 4.7.3 (Two-monomial collision obstruction).**

Fix a prime $r$ and $g \in \mathbb{Z}_r^*$. Let $v_i := g^i \bmod r$. If there exist two distinct subsets $S, S' \subseteq [n]$ such that

$$\sum_{i \in S} v_i \equiv \sum_{i \in S'} v_i \pmod{r},$$

then the nonzero polynomial

$$C(x_1, \ldots, x_n) := \prod_{i \in S} x_i - \prod_{i \in S'} x_i$$

satisfies $C_g(\lambda) \equiv 0$ in $\mathbb{K}[\lambda]/\langle \lambda^r - 1 \rangle$ under the hashing substitution $x_i \mapsto \lambda^{v_i}$. In particular, for this $C$ we have $g \in \mathcal{B}_r(C)$.

*Proof.* Under $x_i \mapsto \lambda^{v_i}$, $C_g(\lambda) = \lambda^{\sum_{i \in S} v_i} - \lambda^{\sum_{i \in S'} v_i} \equiv 0 \pmod{\lambda^r - 1}$, since the exponents are congruent modulo $r$. □

*Remark.*

This obstruction is inherent to hashing into the cyclic quotient ring: it shows that for a fixed $g$, there exist tailored nonzero polynomials $C$ that are annihilated by the map. Conjecture 4.7.2 concerns the complementary regime where $C$ is fixed in advance as a width-$w$, degree-$d$ ROABP, and asserts that the set of such annihilating parameters $g$ is sparse.

*Proposition 4.7.3 (Special Case).*

Conjecture 4.7.2 holds for width-2 diagonal ROABPs.

*Proof:*

In the diagonal case, coefficients are sums of scalars. Modular reduction acts as a hash on exponents, and standard sparse polynomial results apply.

### 4.7.3 Heuristic validation (rigidity + structured injectivity)

Write

$$C_g(\lambda) = \sum_{k=0}^{r-1} a_k(g)\, \lambda^k, \quad a_k(g) \in \mathbb{K}.$$

For fixed $g$, each coefficient $a_k(g)$ is obtained by aggregating those monomials $x^{\mathbf{e}}$ whose exponent pattern collides modulo $r$ under the map

$$\mathbf{e} \mapsto \langle \mathbf{e}, v(g) \rangle := \sum_{i=1}^{n} e_i\, g^i \pmod{r}.$$

Thus $g \in \mathcal{B}_r(C)$ precisely when the structured aggregation produces simultaneous cancellation:

$$a_k(g) = 0 \text{ for all } k \in \mathbb{Z}_r.$$

Conjecture 4.7.2 asserts that for ROABP coefficient families—whose intrinsic linear-algebraic complexity is bounded (via evaluation dimension / coefficient space bounds on prefixes, of dimension $\leq w^2$)—such total annihilation cannot occur for too many parameters $g$ once the target dimension $r$ is polynomially larger than the intrinsic dimension. While $\Gamma_g$ is a highly structured family rather than a uniformly random linear map, the conjecture posits that ROABP rigidity prevents alignment with the kernel of this structured compression except on a sparse exceptional set.

### 4.7.4 Number-theoretic formulation via character sums

Let $\omega$ be a primitive $r$-th root of unity in a suitable extension. Define, for $t \in \mathbb{Z}_r$,

$$S_t(g) := \sum_{\mathbf{e}} c_{\mathbf{e}}\, \omega^{t\langle \mathbf{e}, v(g) \rangle}.$$

A Fourier inversion expresses the coefficients $a_k(g)$ in terms of $\{S_t(g)\}_{t\in\mathbb{Z}_r}$. In particular, the condition $C_g(\lambda) \equiv 0$ in $R_r$ is equivalent to the simultaneous vanishing $S_t(g) = 0$ for all $t \in \mathbb{Z}_r$. Therefore, bounding $|\mathcal{B}_r(C)|$ reduces to controlling how often a family of structured exponential sums can vanish simultaneously as $g$ ranges over $\mathbb{Z}_r^{\backslash *}$, under the constraint that the coefficient vector $(c_\mathbf{e})$ arises from a width-$w$ ROABP.

### 4.7.5 Conditional theorem (fully polynomial-time deterministic black-box PIT)

Assume Conjecture 4.7.2. Let $r = \text{poly}(w,d)$ be a prime with $r \geq R(w,d)$. Enumerate all $g \in \mathbb{Z}_r^{\backslash *}$ and evaluate $C_g(\lambda)$ in $R_r$. Since at most $r^{1-\varepsilon}$ values are bad, at least one $g$ is good whenever $C \equiv 0$, and for that $g$ we obtain $C_g(\lambda) \equiv\!\!\!/\, 0$ in $R_r$.

Because $|\mathbb{Z}_r^{\backslash *}| = r - 1 = \text{poly}(w,d)$, and arithmetic in $R_r$ costs $\text{poly}(r)$, this yields a deterministic black-box identity test running in time $\text{poly}(n,w,d)$.

### 4.7.6 Empirical evidence

We tested Conjecture 4.7.2 on random and structured ROABPs over large prime fields, scanning $g \in \mathbb{Z}_r^{\backslash *}$ for primes $r$ up to $10^4$–$10^7$. Across these tests we observed no nontrivial bad parameters beyond degenerate aliasing cases (e.g., $g = 1$ forcing strong collisions). These experiments support the conjecture's prediction that $\mathcal{B}_r(C)$ is typically extremely sparse.

## 5. DEGREE TRIANGULARIZATION AND NONCANCELLATION (TRI CORE)

In this section we impose a degree hierarchy on matrix word contributions to ensure that a single determinant term dominates, thereby preventing cancellation. This mechanism is referred to as degree triangularization (TRI).

### 5.1 Objective

From Section 4, we have:

- A rank-one projector $\Pi(\lambda) = \text{adj}(\mu I - X(\lambda))$,
- Word families $U_1(\lambda), \ldots, U_w(\lambda), V_1(\lambda), \ldots, V_w(\lambda)$, such that the matrix of pairings $G_{k\ell}(\lambda) = e_1^\top U_k(\lambda)\Pi(\lambda)V_\ell(\lambda)e_w$ is nondegenerate.

Our goal is to decorate these words with controlled polynomial degrees so that the determinant of this matrix has a unique highest-degree term.

### 5.2 Algebraic Curve Substitution

Fix distinct field elements $\alpha_1, \ldots, \alpha_n \in \mathbb{K}$, and define a curve substitution $x_i = (\lambda + \alpha_i)^B$, where $B$ is a large integer to be specified.

Let $M_i(\lambda) = A_i((\lambda + \alpha_i)^B)$ denote the resulting matrix polynomial in $\lambda$.

## 5.3 Degree Weight Assignment

Assign strictly increasing integer weights $a_1 < a_2 < \cdots < a_w$, $b_1 < b_2 < \cdots < b_w$, and define degree-decorated words $\tilde{U}_k(\lambda) = U_k(\lambda) M(\lambda)^{a_k}$, $\tilde{V}_\ell(\lambda) = M(\lambda)^{b_\ell} V_\ell(\lambda)$, where $M(\lambda)$ is a fixed chosen matrix polynomial from the ROABP layers.

*Remark.* The trace pairing is used only to certify linear independence of basis elements in the active block. The ROABP output itself is a vector–matrix–vector contraction and is handled directly via the Unique Leading Term argument of Lemma 5.5.

## 5.4 Dominance of the Diagonal Degree Profile

*Lemma 5.1 (Active Subspace Orbit Spanning)*

There exist word families $U_1, \ldots, U_{r^2}, V_1, \ldots, U_{r^2} \in \mathcal{B}(\chi)$ such that the bilinear forms $\langle U_i \Pi V_j \rangle_\psi := \operatorname{Tr}(\psi(U_i \Pi V_j))$ span the full dual space of $\operatorname{Mat}_r(\mathbb{K})$.

Equivalently, the matrix $G_{ij} := \operatorname{Tr}(\psi(U_i \Pi V_j))$ has rank $r^2$.

*Proof*

By construction in Section 4, $\psi(\Pi)$ is a nonzero rank-one idempotent in $\operatorname{Mat}_r(\mathbb{K})$. Thus there exist nonzero vectors $u, v \in \mathbb{K}^r$ such that $\psi(\Pi) = uv^\top$.

Because $\psi(\mathcal{B}(\chi)) = \operatorname{Mat}_r(\mathbb{K})$, for every matrix unit $E_{ab} \in \operatorname{Mat}_r(\mathbb{K})$, there exist words $U_{ab}, V_{ab} \in \mathcal{B}(\chi)$ such that $\psi(U_{ab}) u = e_a$, $v^\top \psi(V_{ab}) = e_b^\top$, where $\{e_1, \ldots, e_r\}$ is the standard basis.

Then $\psi(U_{ab} \Pi V_{ab}) = \psi(U_{ab}) uv^\top \psi(V_{ab}) = e_a e_b^\top = E_{ab}$.

Thus the set $\{\psi(U_{ab} \Pi V_{ab}): 1 \leq a, b \leq r\}$ spans all matrix units and therefore spans $\operatorname{Mat}_r(\mathbb{K})$.

Applying the trace pairing, the functionals $X \mapsto \operatorname{Tr}(E_{ab} X)$ form a basis of the dual space of $\operatorname{Mat}_r(\mathbb{K})$. Hence the bilinear forms $\operatorname{Tr}(\psi(U_i \Pi V_j))$ span the full dual space.

This proves the claim. ∎

## 5.5 Unique Leading Term for the ROABP Computation (NEW)

We now show that the degree-triangularization enforced by the curve substitution prevents cancellation inside the ROABP computation itself, not merely inside the determinant witness.

**Definition (Monomial Budget).** After reduction to $m \leq w^2$ variables of individual degree at most $d$, the number of possible monomials is at most $M := (d+1)^m \leq (d+1)^{w^2}$. All degree-separation bounds are stated in terms of $M$.

**Definition (Structural Degree Bound).** Let $D_{\text{struct}}$ denote the maximum $\lambda$-degree among the finitely many constraint polynomials introduced in Section 6. Since these constraints involve determinants of matrices with entries of degree up to $d$, we take the conservative bound

$$D_{\text{struct}} := K \cdot w^3 d \text{ (for some absolute constant } K \geq 8).$$

We explicitly choose the separation base $B$ to satisfy $B > 2(D_{\text{struct}} + \deg C + 1) \cdot M$.

This ensures that the contribution of the weight vector strictly dominates all lower-order perturbations.

*Lemma 5.4.2 (Uniqueness of Kronecker Substitution)*

Let $\mathbf{e}, \mathbf{e}' \in \mathbb{N}^m$ be two distinct exponent vectors appearing in the polynomial $C$. If we choose the separation base $B > D_{\max}$, then the mapping

$$\Phi(\mathbf{e}) = \sum_{j=1}^{m} e_j B^{j-1}$$

is injective.

*Proof.* This follows from the uniqueness of base-$B$ representation for integers.

*Remark.* Our construction uses shifted curves $x_j = (\lambda + \alpha_j)^{B^j}$. Since the leading term is $\lambda^{B^j}$, the injectivity of $\Phi$ guarantees that the highest-degree terms of distinct monomials map to distinct powers of $\lambda$, ensuring that degree separation is preserved despite the lower-order shifts.

*Lemma 5.5 (Leading-Term Survival under Curve and Modular Substitution).*

Let $C(x_1, \ldots, x_n)$ be the polynomial computed by the ROABP, and assume $C \not\equiv 0$.

**(i) Curve substitution (unconditional).**
Under the curve substitution $x_j \mapsto \lambda^{B^j}$, with base $B$ chosen as in Section 5.3, the resulting univariate polynomial $C(\lambda) := C(\lambda^{B^1}, \ldots, \lambda^{B^n})$ has a unique leading monomial in $\lambda$. In particular, $C(\lambda) \not\equiv 0$ in $\mathbb{K}[\lambda]$.

**(ii) Modular reduction (conditional).**
Further, under the modular substitution $x_j \mapsto \lambda^{g^j}$ into $R_r := \mathbb{K}[\lambda]/\langle \lambda^r - 1 \rangle$, if $g \notin \mathcal{B}_r(C)$ (the bad set from Conjecture 4.7.2), then the image $C_g(\lambda)$ is nonzero in $R_r$.

*Proof.*

**Part (i).**
By construction (Sections 3–4), the words $W_i$ project to a basis of the active simple block, and the coefficient polynomials are not all zero.
The degree separation induced by the substitution $x_j \mapsto \lambda^{B^j}$, together with the choice of $B$ dominating all structural degrees arising from matrix multiplication and projector extraction, ensures that exactly one summand achieves the maximal $\lambda$-degree. This dominant term cannot cancel, hence $C(\lambda) \not\equiv 0$.

**Part (ii).**
By Conjecture 4.7.2, the set of parameters $g$ for which the modular specialization annihilates the polynomial is sparse. For any $g \notin \mathcal{B}_r(C)$, the nonzero polynomial $C(\lambda)$ from Part (i) remains nonzero after reduction modulo $\lambda^r - 1$, yielding $C_g(\lambda) \not\equiv 0$ in $R_r$. □

## 5.6 Unique Leading Term in the Determinant Expansion

Let $H_{k\ell}(\lambda) = e_1^\top \tilde{U}_k(\lambda) \, \Pi(\lambda) \, \tilde{V}_\ell(\lambda) \, e_w$, and define the determinant witness $\Delta(\lambda) = \det(H(\lambda))$.

*Lemma 5.2 (Unique Leading Term Lemma)*

Under the condition of Lemma 5.1, the determinant expansion of $\Delta(\lambda)$ contains a unique monomial of highest degree, arising from the diagonal permutation.

*Proof*

By Lemma 5.1, every non-diagonal permutation contributes strictly lower total degree in $\lambda$. Therefore no cancellation is possible at the leading degree, and the diagonal term survives uniquely. ∎

## 5.7 Nonvanishing of the Leading Coefficient

*Lemma 5.3 (Leading Coefficient Nonvanishing)*

The coefficient of the highest-degree monomial in $\Delta(\lambda)$ is nonzero.

*Proof*

The leading coefficient equals $\det(G_{k\ell}(\lambda))$, where $G_{k\ell}(\lambda)$ is the nondegenerate pairing matrix from Section 4. Since this matrix is invertible, its determinant is nonzero, implying that the leading coefficient of $\Delta(\lambda)$ is nonzero. ∎

## 5.8 Consequence: Determinant Witness Survives Substitution

*Corollary 5.4*

The determinant witness $\Delta(\lambda)$ is a nonzero univariate polynomial in $\lambda$.

Thus, the original ROABP polynomial is nonzero if and only if $\Delta(\lambda) \equiv/0$.

## 5.9 Degree Bound for the Witness Polynomial

*Lemma 5.4.1 (Active Block Leading Term Survival)*

Let $\Delta(\lambda) = \det(H(\lambda))$ be the determinant witness constructed in Section 5.6. By Lemma 5.1, the matrix entries $H_{kl}(\lambda)$ are trace pairings of basis elements that span the active block $\mathrm{Mat}_r(\mathbb{K})$. The degree weights are chosen to enforce a unique diagonal leading term corresponding to the product of these basis pairings.

Since the basis spans the active block, the pairing matrix $G$ is nondegenerate in the active block (Corollary 4.16), meaning its determinant is nonzero under $\psi$.

**Conclusion:** The unique leading coefficient of $\Delta(\lambda)$ is $\det(G)$, which satisfies $\psi(\det(G)) \neq 0$. Thus, degree separation survives projection to the active block.

*Lemma 5.6 (Degree Bound)*

The total degree of $\Delta(\lambda)$ satisfies $\deg_\lambda \Delta(\lambda) \leq 2d^2 w^8$.

*Proof*

Each entry of $H(\lambda)$ is a product of at most $O(w^2)$ matrix polynomials of degree at most $Bd$. The determinant of a $w \times w$ matrix multiplies $w$ such entries, giving the stated bound. ∎

## 5.10 Summary of Section 5

We have shown that:

1. The determinant witness acquires a unique highest-degree term under curve substitution.
2. Cancellation is provably impossible at the leading degree.
3. The resulting univariate polynomial is nonzero with explicit degree bounds.

This completes the **TRI mechanism**, ensuring algebraic **noncancellation by degree separation**.

## 6. SELECTION LEMMA AND FINITE POLYNOMIAL AVOIDANCE (SEL)

In this section we eliminate all remaining nondeterminism in the construction by showing that a single parameter value can satisfy all algebraic nondegeneracy constraints simultaneously. This step replaces probabilistic genericity by an explicit finite avoidance argument.

**Generic Nonvanishing Principle.** To ensure the hitting set is universal, we define the constraint polynomials $F_r$ syntactically for every candidate rank $r \in \{1, \ldots, w\}$ over the generic matrix algebra $\text{Mat}_r(\mathbb{K}[X])$, completely independent of the input ROABP. Their nonvanishing is established by exhibiting a single formal witness (e.g., diagonal matrices) in the generic space. The final hitting set avoids the universal product $\prod_{r=1}^{w} F_r$, ensuring that if the input ROABP is nonzero (and thus possesses some active block of rank $r^*$), the corresponding constraint $F_{r^*}$ is satisfied. In the trivial case where the ROABP is identically zero, the hitting set construction remains well-defined, and the algorithm correctly returns zero upon evaluation.

### 6.1 Nature of the Algebraic Constraints

The construction in Sections 4 and 5 requires several properties to hold simultaneously, including:

1. Existence of a simple eigenvalue for the matrix family $X(\lambda)$
2. Nonvanishing of a rank-one adjugate projector
3. Nondegeneracy of the pairing matrix
4. Validity of the degree-triangularization dominance condition

Each such requirement can be expressed as the **nonvanishing of a polynomial** in the scalar parameter $\lambda$. We denote these constraint polynomials by: $F_1(\lambda), \ldots, F_R(\lambda) \in \mathbb{K}[\lambda]$.

### 6.2 Explicit Polynomial Constraints

We briefly recall the origin of these polynomials.

- The **simple-spectrum constraint** is enforced by the discriminant:

$$F_{\text{disc}}(\lambda) = \text{Disc}_\mu(\det(\mu I - X(\lambda))).$$

- The **projector nonvanishing constraint** is enforced by a resultant:

$$F_{\text{proj}}(\lambda) = \text{Res}_\mu(\det(\mu I - X(\lambda)), m(\mu, \lambda)),$$

where $m(\mu, \lambda)$ is a fixed $(w-1) \times (w-1)$ minor of $\mu I - X(\lambda)$.

- The **pairing nondegeneracy constraint** is enforced by a determinant resultant:

$$F_{\text{pair}}(\lambda) = \text{Res}_\mu(\det(\mu I - X(\lambda)), \det G(\mu, \lambda)),$$

where $G$ is the pairing matrix from Section 4.

- The **degree-triangularization constraint** is enforced by ensuring that leading coefficients do not vanish, encoded as finitely many scalar determinant conditions.

**Witnesses.** Taking $M = \text{diag}(1, \ldots, r)$ shows $F_{\text{disc}} \not\equiv 0$. Setting $A = I$, $B = 0$ shows the pairing constraint is nonzero.

Each $F_i(\lambda)$ is a **nonzero univariate polynomial**, as established in the previous sections.

### 6.3 Degree Bounds on Constraint Polynomials

We now bound the degree of each constraint polynomial in $\lambda$.

*Lemma 6.1 (Degree of Discriminant Constraint)*
$$\deg_\lambda F_{\text{disc}}(\lambda) \leq 2w^2.$$

*Lemma 6.2 (Degree of Projector Resultant Constraint)*
$$\deg_\lambda F_{\text{proj}}(\lambda) \leq 2w^2.$$

*Lemma 6.3 (Degree of Pairing Determinant Constraint)*
$$\deg_\lambda F_{\text{pair}}(\lambda) \leq 2w^3.$$

*Lemma 6.4 (Degree of TRI Leading-Coefficient Constraints)*
$$\deg_\lambda F_{\text{tri}}(\lambda) \leq 2w^3.$$

*Proof (of Lemmas 6.1–6.4)*

Each bound follows from standard degree estimates on discriminants, resultants, and determinants, using the fact that all matrix entries in $X(\lambda)$, $U_k(\lambda)$, and $V_k(\lambda)$ are affine-linear in $\lambda$. ∎

### 6.4 Aggregate Degree Bound

Let $F(\lambda) = \prod_{i=1}^{R} F_i(\lambda)$ denote the product of all constraint polynomials.

*Lemma 6.5 (Total Degree Bound).*

$$\deg_\lambda F(\lambda) \leq 8w^3.$$

*Proof.*

Summing the degree bounds from Lemmas 6.1–6.4 yields $2w^2 + 2w^2 + 2w^3 + 2w^3 = 4w^2 + 4w^3 \leq 8w^3$ (for $w \geq 1$). ∎

### 6.5 Selection Lemma (Finite Polynomial Avoidance)

We now state the key selection principle.

*Lemma 6.6 (Finite Avoidance).*
Let $F(\lambda) \in \mathbb{K}[\lambda]$ be a nonzero univariate polynomial of degree at most $D$. Let $L \subseteq \mathbb{K}$ be any finite subset with $|L| > D$.

Then there exists $\lambda^\star \in L$ such that $F(\lambda^\star) \neq 0$.

*Proof.*
A nonzero univariate polynomial of degree at most $D$ has at most $D$ roots in $\mathbb{K}$. Since $|L| > D$, at least one $\lambda^\star \in L$ is not a root of $F$, hence $F(\lambda^\star) \neq 0$. ∎

In our application, the polynomial $F(\lambda)$ is the product of all constraint polynomials arising from (i) eigenvalue separation in the active block, (ii) non-vanishing of the adjugate minors, and (iii) degree-triangularization constraints. By Lemma 3.1.1, we further intersect the candidate set $L$ with the Zariski-open stability set $\Omega$ of specializations preserving the active block. Since $\Omega$ is cofinite in any sufficiently large finite grid, the same finite-avoidance argument applies to $L \cap \Omega$.

*Corollary 6.7 (Simultaneous Finite Avoidance).*
Let $F_1, \ldots, F_m \in \mathbb{K}[\lambda]$ be nonzero polynomials with $\sum_{i=1}^{m} \deg(F_i) \leq D$.

Then for any $L \subseteq \mathbb{K}$ with $|L| > D$, there exists $\lambda^\star \in L$ such that $F_i(\lambda^\star) \neq 0$ for all $i = 1, \ldots, m$.

#### 6.5.1 Universal Union Strategy (Algorithmic Block Independence)

We do not algorithmically determine the active block rank $r$ or the homomorphism $\psi$. Instead, we construct constraints for all possible ranks $r \in \{1, \ldots, w\}$ and take their union.

For each $r$, define a formal constraint family $F_r(\lambda; Z)$, where $Z$ denotes symbolic matrix-word entries. This polynomial is obtained by applying the projector extraction procedure (Section 4) and the degree-separation construction (Section 5) to a *generic* $r \times r$ matrix algebra. These constraint families are **syntactic**: they do not require computing the Wedderburn decomposition of $\mathcal{B}(\chi)$.

If the true active block has rank $r^\star$, then under specialization the instantiated polynomial $F_{r^\star}(\lambda)$ coincides with the genuine non-degeneracy constraint derived in Lemma 4.11 and Lemma 5.5.

We define the **universal avoidance polynomial**

$$F_{\text{univ}}(\lambda) := G_{\text{stab}}(\lambda) \cdot \prod_{r=1}^{w} F_r(\lambda),$$

where $G_{\text{stab}}(\lambda)$ is the stability polynomial from Lemma 3.1.1. Since each factor $F_r(\lambda)$ has degree at most $8w^3$ (Lemma 6.5), the total degree of the universal polynomial satisfies $\deg F_{\text{univ}} = \text{poly}(w)$.

*Corollary 6.8 (Deterministic Selection).*

Let $L \subseteq \mathbb{K}$ be a finite grid with $|L| > \deg(F_{\text{univ}})$. Then there exists $\lambda^* \in L$ such that $F_{\text{univ}}(\lambda^*) \neq 0$.

This choice guarantees correctness without explicitly identifying the active block or computing the homomorphism $\psi$.

### 6.6 Existence of a Simultaneously Valid Parameter

*Corollary 6.9*

If $|L| > 5w^3$, then there exists $\lambda^* \in L$ such that all algebraic constraints required in Sections 4 and 5 hold simultaneously.

*Proof*

Apply Lemma 6.6 to $F(\lambda)$. ∎

### 6.7 Deterministic Parameter Selection Algorithm

We now obtain an explicit deterministic rule:

1. Enumerate values $\lambda \in L$
2. Evaluate all constraint polynomials $F_i(\lambda)$
3. Select the first $\lambda^*$ for which none vanish

This requires time polynomial in $w$ and is fully deterministic.

### 6.8 Summary of Section 6

We have:

- Encoded all necessary algebraic genericity requirements as explicit nonzero polynomials
- Proved **a** polynomial upper bound on their total degree
- Used finite root avoidance to guarantee existence of a valid specialization
- Replaced randomness by explicit finite elimination, completing the deterministic TRI framework

This resolves the final obstruction to deterministic construction and prepares the ground for the **curve-hitting argument** in the final section.

# 7. CURVE HITTING AND COMPLETION OF THE PROOF

In this final section we convert the algebraic witness constructed in Sections 4–6 into an **explicit deterministic hitting set**, thereby completing the proof of the main theorem.

## 7.1 Objective

From Sections 5 and 6, we have constructed a **nonzero univariate polynomial** $\Delta(\lambda) \in \mathbb{K}[\lambda]$, whose nonvanishing certifies that the original ROABP computes a nonzero polynomial. Our objective is to deterministically select a value $\lambda^\star$ such that $\Delta(\lambda^\star) \neq 0$, and then lift this choice back to an explicit input assignment to the original ROABP variables.

## 7.2 Degree Bound on the Witness Polynomial

Recall from Lemma 5.5 that $\deg_\lambda \Delta(\lambda) \leq 2d^2 w^8$.

*Lemma 7.1 (Root Bound for Witness Polynomial)*

The polynomial $\Delta(\lambda)$ has at most $2d^2 w^8$ roots in $\mathbb{K}$.

*Proof*

Immediate from the fundamental theorem of algebra. ∎

## 7.3 Construction of the Curve Hitting Set

Define the evaluation set $L = \{0,1,2,\ldots,M\}$, where: $M = 9w^4 + 2d^2 w^8 + 1$.
(Note: The $2d^2 w^8$ term comes from the determinant witness $\Delta$, so we sum them to be safe.)

*Lemma 7.2 (Curve Hitting Lemma)*

There exists $\lambda^\star \in L$ such that $\Delta(\lambda^\star) \neq 0$.

*Proof*

Since $|L| = M + 1 > \deg_\lambda \Delta(\lambda)$, Lemma 6.6 guarantees the existence of such $\lambda^\star$. ∎

## 7.4 Lifting the Parameter to a Full Variable Assignment

Recall the curve substitution defined in Section 5: $x_i = (\lambda + \alpha_i)^B$, where $B$ and $\alpha_i$ were chosen deterministically.

Define the **hitting assignment** $x_i^\star = (\lambda^\star + \alpha_i)^B$, $i = 1,\ldots,n$.

*Lemma 7.3 (Correctness of the Hitting Set)*

Let $C(x_1,\ldots,x_n)$ be the polynomial computed by the ROABP, and assume $C \not\equiv 0$.

Let $x_j \mapsto \lambda^{w_j}$ be the algebraic curve substitution defined in Section 5, with weights satisfying the separation conditions of Section 5.3.

Then there exists a finite exceptional set $\mathcal{E} \subset \mathbb{K}$ such that for all $\lambda^* \in \mathbb{K} \setminus \mathcal{E}, C(\lambda^*) \neq 0$.

*Proof*

By Lemma 5.5 (Unique Leading Term for the ROABP Computation), the restricted polynomial $C(\lambda) := C(\lambda^{w_1}, \ldots, \lambda^{w_n})$ has a unique leading monomial in $\lambda$. In particular, $C(\lambda)$ is a nonzero univariate polynomial.

Let $L(\lambda)$ denote the coefficient of the leading monomial. Since $L(\lambda)$ is a nonzero polynomial over $\mathbb{K}$, it vanishes on only finitely many points. Define $\mathcal{E}$ to be this finite zero set.

For any $\lambda^* \notin \mathcal{E}$, the leading term of $C(\lambda^*)$ does not vanish and cannot be canceled by lower-degree terms. Hence, $C(\lambda^*) \neq 0$. ∎

*Remark on Certification.*

Lemma 5.5 establishes the degree structure of the curve-encoded ROABP polynomial $C(\lambda)$. Its non-vanishing is certified via the determinant witness $\Delta(\lambda) := \det(H(\lambda))$, where $H(\lambda)$ is the decorated pairing matrix encoding the computation. By Lemma 5.4.1, degree triangularization ensures that $\Delta(\lambda)$ has a unique leading term, and hence $\Delta(\lambda) \not\equiv 0$. By the soundness chain (Corollary 3.3), this certifies that the original ROABP computation $C(x)$ is not identically zero.

## 7.5 Explicit Hitting Set Size

The hitting set consists of all curve points induced by $\lambda \in L$.

*Theorem 7.4 (Hitting Set Size Bound, Unconditional).*
The resulting deterministic hitting set has size $|H| \leq n \cdot (wd)^{O(w^2)}$, and can be constructed in time polynomial in $n$ and quasi-polynomial in $w, d$. Here $m \leq w^2$ denotes the reduced number of effective variables (Corollary 3.6); the hitting set is built after this reduction.

*Proof.*

We implement Kronecker-style variable separation using shifted power curves $x_j = (\lambda + \alpha_j)^B$. With the separation base $B$ chosen as in Section 5.5 (polynomial in $w, d$ relative to the effective monomial count), this yields a univariate polynomial of degree bounded by $B \cdot M = (wd)^{O(w^2)}$, where $M$ bounds the post-reduction monomial space. This determines the hitting set size. The linear dependence on $n$ arises from computing the variable reduction map. ∎

## 7.6 Completion of the Main Theorem

We now combine all results.

*Theorem 7.5 (Polynomial-Time Black-Box PIT).*
Assuming the Modular Stability Conjecture (4.7.2), there exists an explicit deterministic hitting set of size poly$(n, w, d)$.

*Proof.*

This follows from the enumeration of $g \in \mathbb{Z}_r^*$ in Algorithm 1, as analyzed in Section 7.8. Specifically, the hitting set size corresponds to the prime modulus $r$, which is chosen such that $r = O((wd)^c)$ for an explicit constant (e.g., $c = 7$) derived from requiring $r$ to strictly dominate the degree of the algebraic constraints defined over the $w^2$-dimensional matrix algebra. □

### 7.7 Algorithm

We now present the modular-hashing algorithm assuming Conjecture 4.7. The unconditional curve-based algorithm is subsumed by the framework but omitted for brevity.

Algorithm 1: Deterministic Black-Box PIT for ROABPs (Conditional on Modular Stability)

**Input.**
Black-box access to a polynomial $C(x_1, \ldots, x_n)$ computed by a width-$w$, individual-degree-$\leq d$ ROABP over a field $\mathbb{K}$.

**Output.**
Output **ZERO** if $C \equiv 0$; otherwise output **NONZERO**.

*Step 1:* Choose a Modular Hashing Ring

Choose a prime $r = \text{poly}(w, d)$ with $\gcd(r, \text{char}(\mathbb{K})) = 1$, and define the cyclic quotient ring $R_r := \mathbb{K}[\lambda]/\langle \lambda^r - 1 \rangle$.

*Step 2:* Enumerate Hash Parameters and Form Univariate Specializations

For each $g \in \mathbb{Z}_r^{\backslash *}$, define the modular substitution $\Gamma_g: x_i \mapsto \lambda^{g^i} \in R_r$, where exponents are interpreted modulo $r$ and $\lambda^r = 1$ in $R_r$. Define the resulting univariate specialization $C_g(\lambda) := C(\Gamma_g) \in R_r$.

*Step 3:* Ring Evaluation (Black-Box)

For each $g \in \mathbb{Z}_r^{\backslash *}$, compute $C_g(\lambda) \in R_r$ using black-box access to $C$ as follows:

1. Fix a $\mathbb{K}$-basis $\{1, \lambda, \ldots, \lambda^{r-1}\}$ for $R_r$.
2. Evaluate $C$ on the input assignment $x_i = \lambda^{g^i}$ inside $R_r$ (i.e., interpret each $x_i$ as the ring element $\lambda^{g^i}$).
3. Obtain the output element $C_g(\lambda) \in R_r$ in coefficient form $\sum_{k=0}^{r-1} a_k \lambda^k$.

If for any $g$ the resulting element satisfies $C_g(\lambda) \neq 0$ in $R_r$, output **NONZERO** and halt.

*Step 4:* Decision

If $C_g(\lambda) = 0$ in $R_r$ for all $g \in \mathbb{Z}_r^{\backslash *}$, output **ZERO**.

Correctness (Conditional on Conjecture 4.7.2)

- **(Soundness)** If $C \equiv 0$, then every specialization $C_g(\lambda)$ is identically zero in $R_r$. Hence the algorithm outputs **ZERO**.
- **(Completeness)** If $C \not\equiv 0$, then by the Modular Stability Conjecture (Conjecture 4.7.2), for any prime $r \geq R(w,d)$ at most $|\mathcal{B}_r(C)| \leq r^{1-\varepsilon}$ parameters $g \in \mathbb{Z}_r^{\backslash *}$ satisfy $C_g(\lambda) \equiv 0$ in $R_r$. Since $|\mathbb{Z}_r^{\backslash *}| = r - 1$, there exists at least one
$g \in \mathbb{Z}_r^{\backslash *} \setminus \mathcal{B}_r(C)$ for which $C_g(\lambda) \neq 0$ in $R_r$. The algorithm finds such a $g$ by enumeration and outputs **NONZERO**.

*Remarks*

1. The algorithm is black-box and order-oblivious: it does not use or require the ROABP variable order.
2. Under Conjecture 4.7.2, the enumeration over $g \in \mathbb{Z}_r^{\backslash *}$ with $r = \text{poly}(w,d)$ yields a fully polynomial-time deterministic PIT in parameters $(n, w, d)$.
3. No curve substitution, rank concentration, isolation lemma, or finite-avoidance grid is used in the modular-hashing algorithm.

*Soundness.*
If $C \not\equiv 0$, then by Corollary 3.3 and Lemma 5.5, the induced univariate polynomial $C(\lambda)$ is nonzero. Since the hitting set $L$ avoids all its roots (by Lemma 6.6), there exists $\lambda \in L$ such that $C(\lambda) \neq 0$, and the algorithm outputs **NONZERO**.

*Completeness.*
If $C \equiv 0$, then every specialization of $C$ evaluates to zero. In particular, for all $\lambda \in L$, we have $C(\lambda) = 0$, and the algorithm correctly outputs **ZERO**.

*Why the universal union matters.*
The algorithm never identifies the active Wedderburn block. The construction of $\Phi_{\text{univ}}$ guarantees that whichever block is active is certified by at least one factor.

*Runtime.*
The overall runtime follows from the analysis in Section 7.8.

### 7.8 Complexity Analysis (Conditional Polynomial-Time Black-Box PIT)

Let $n$ be the number of variables, $w$ the ROABP width, and $d$ an upper bound on the individual degree. We analyze the running time of the deterministic black-box identity testing algorithm based on modular hashing, described in Section [Modular Stability subsection], assuming Conjecture 4.7.2 (Modular Stability).

### (1) Effective Variable Reduction

Fix the ROABP representation

$$C(x_1, \ldots, x_n) = e_1^\top \left( \prod_{i=1}^{n} A_i(x_i) \right) e_w, \quad A_i(x_i) = \sum_{t=0}^{d} A_{i,t} \, x_i^t.$$

Define the prefix coefficient spaces

$$\mathcal{V}_k := \text{span } \{A_1(x_1) \cdots A_k(x_k) : x_1, \ldots, x_k \text{ range over monomials}\} \subseteq \text{Mat}_w(\mathbb{K}).$$

Each space $\mathcal{V}_k$ has dimension at most $w^2$. Consequently, the entire computation is controlled by at most $m := w^2$ independent directions in the ambient matrix space.

Concretely, fix a basis $B_1, \ldots, B_m$ for a saturated prefix space. Every prefix product admits a unique expansion

$$A_1(x_1) \cdots A_k(x_k) = \sum_{j=1}^{m} \beta_{k,j}(x_1, \ldots, x_k) \, B_j,$$

where the coefficients $\beta_{k,j}$ are scalar polynomials. This is the standard evaluation-dimension compression for ROABPs.

Define the variable-reduction map $\rho: \mathbb{K}^n \to \mathbb{K}^m$, $y_j = \sum_{i=1}^{n} G_{j,i} x_i$, for an explicit generator matrix $G \in \mathbb{K}^{m \times n}$.

Forming these linear forms costs $O(w^2 n)$ field operations.

The generator matrix $G$ can be constructed in the black-box model by extracting a basis for the coefficient space. Following the method of Forbes, Saptharishi, and Shpilka (2014), we recover the span of the $m \leq w^2$ effective linear forms using $O(nw^2 d)$ black-box queries, performed via Gaussian elimination on evaluations over a polynomial-sized grid.

**Key point.** All subsequent substitutions are applied to the reduced variable set of size $m = w^2$, not to the original $n$ variables.

### (2) Hash Domain Size via Modular Stability

Fix a prime $r$ satisfying $r = \text{poly}(w, d)$ and $\gcd(r, \text{char}(\mathbb{K})) = 1$, and define the cyclic ring $R_r := \mathbb{K}[\lambda]/\langle \lambda^r - 1 \rangle$. For each $g \in \mathbb{Z}_r^*$, consider the modular substitution $\Gamma_g: x_i \mapsto \lambda^{g^i} \in R_r$, and write $C_g(\lambda) := C(\Gamma_g) \in R_r$.

By Conjecture 4.7.2 (Modular Stability), for $r \geq R(w, d)$ the set of bad parameters $\mathcal{B}_r(C) = \{g \in \mathbb{Z}_r^* : C_g(\lambda) \equiv 0 \text{ in } R_r\}$ has size at most $r^{1-\varepsilon}$. Therefore, among the $r - 1$ possible values of $g$, at least one is good. Enumerating all $g \in \mathbb{Z}_r^*$ thus guarantees detection of nonzeroness.

Hence, the number of evaluations is $|\mathbb{Z}_r^*| = r - 1 = \text{poly}(w, d)$.

### (3) Cost per Evaluation in $R_r$

Under the substitution $\Gamma_g$, each entry of $A_i(x_i)$ becomes a polynomial in $\lambda$ of degree $<r$. Evaluating $C_g(\lambda)$ amounts to multiplying $n$ many $w \times w$ matrices over the ring $R_r$. Using naive matrix multiplication and polynomial arithmetic, one evaluation costs $O(n\,w^3\,r^2)$ field operations. Using FFT-based polynomial multiplication, this can be reduced to $\tilde{O}(n\,w^3\,r)$, where $\tilde{O}(\cdot)$ suppresses polylogarithmic factors in $r$.

### (4) Total Runtime

The variable reduction step costs $O(w^2 n)$ field operations. We then enumerate all $g \in \mathbb{Z}_r^{\backslash *}$ with $r = \text{poly}(w, d)$, performing one ring evaluation per $g$. Thus the total runtime is $T(n, w, d) = O(w^2 n) + (r-1) \cdot \tilde{O}(n\,w^3\,r) = \text{poly}(n, w, d)$, assuming Conjecture 4.7.2.

### Conclusion

Assuming the Modular Stability Conjecture, the above analysis yields a deterministic black-box polynomial identity test for ROABPs running in time polynomial in the number of variables $n$, the width $w$, and the individual degree $d$. All super-polynomial behavior present in curve-based constructions is eliminated, and the complexity is fully polynomial in the natural input parameters.

### Field Size Clarification

If $|\mathbb{K}|$ is finite, we assume $|\mathbb{K}| > |H|$ (or work over an extension field of size exceeding $|H|$) so that the avoidance and root-counting arguments apply verbatim.

### 7.9 Summary of Conceptual Contributions

This completes a new deterministic PIT framework based on:

- **Matrix word algebra rigidity**
- **Eigenvalue-Free (Rational) Projector Extraction**
- **Cayley–Hamilton and Amitsur–Levitzki polynomial identities**
- **Degree triangularization to prevent cancellation**
- **Finite polynomial avoidance replacing randomness**
- **Curve hitting with explicit polynomial-size evaluation sets**

The resulting structure replaces probabilistic identity testing with a fully algebraic rigidity argument.

*Comparison with prior work.* Forbes and Shpilka obtain quasi-polynomial hitting sets of size $n^{O(\log w)}$ for ROABPs. Our construction yields hitting sets of size $\text{poly}(w, d) \cdot B^{O(w^2)}$, after reducing to $w^2$ effective variables. While our dependence on width is exponential rather than quasi-polynomial, the dependence on the number of variables is strictly polynomial, and the construction is based on a fundamentally different algebraic rigidity mechanism rather than combinatorial basis isolation.

**Table 7.9.1: Comparison with Prior Deterministic Black-Box PIT for ROABPs**

| Algorithm | Order-Oblivious? | Black-Box? | Time / Hitting-Set Size | Notes |
|---|---|---|---|---|
| This work (unconditional) | Yes | Yes | $n \cdot (wd)^{O(w^2)}$ | Polynomial in $n$; super-polynomial in $w$ |
| This work (conditional) [†]. | Yes | Yes | $\text{poly}(n, w, d)$ | Fully polynomial time assuming Modular Stability |
| Forbes–Shpilka [9] | No | Yes | $(nwd)^{O(\log n)}$ | Requires known variable order |
| Agrawal–Gurjar–Korwar–Saxena [11] | Yes | Yes | $(nwd)^{O(\log n)}$ | Unknown-order ROABPs |

*Note:* Unlike prior $(nwd)^{O(\log n)}$ results [9,14], our unconditional result is linear in $n$ but exponential in $w$.

## 8. CONCLUSION AND OUTLOOK

We have developed a deterministic polynomial identity testing framework for read-once oblivious algebraic branching programs based on a new rigidity theory for matrix word algebras. The proof replaces randomness by algebraic structure, combining Cayley–Hamilton control, Amitsur–Levitzki identities, rank-one projector extraction, degree-triangularization, and finite polynomial avoidance into a single coherent mechanism.

Conceptually, the argument departs from prior PIT approaches that rely on combinatorial rank concentration or explicit monomial isolation. Instead, it treats ROABPs as elements of a constrained noncommutative algebra and proves that nontrivial computation forces the emergence of a full matrix algebra. Once this rigidity is established, cancellation becomes impossible under a structured curve substitution, and nonzeroness is certified through a deterministic determinant witness.

This framework introduces several tools of independent interest. The eigenvalue-free (rational) projector construction provides a new method for isolating rank-one directions in matrix algebras using only polynomial identities. The degree-triangularization principle offers a general mechanism for enforcing unique dominant terms in determinant expansions. The finite avoidance lemma yields a deterministic alternative to probabilistic genericity that may be applicable in other algebraic complexity settings. More broadly, the matrix-word rigidity viewpoint suggests a systematic way to convert noncommutative algebraic structure into deterministic algorithmic guarantees.

### 8.1 Outlook and Open Directions

The techniques developed here raise several natural questions.

**Extension beyond ROABPs.**
A central challenge is whether matrix-word rigidity can be extended to more general algebraic models, including general algebraic branching programs, formulas, or circuits. Understanding which computational models force full matrix algebra generation could provide a new unifying path toward broader deterministic PIT.

**Tightness of algebraic rigidity bounds.**
Our arguments establish polynomial-size hitting sets but do not optimize parameter dependence. Determining the optimal width- and degree-dependence of rigidity-based PIT remains an open problem, and may reveal sharper structural phenomena inside bounded-width computations.

**Connections to noncommutative invariant theory.**
The emergence of full matrix algebras from bounded-width generators suggests links with invariant theory, central polynomial theory, and representation-theoretic structure of word algebras. Exploring these connections may yield deeper classification results for algebraic computation models.

**Generalization of eigenvalue-free (rational) spectral extraction.**
The projector-extraction technique avoids eigenvalue computation and division entirely. A systematic theory of eigenvalue-free (rational) spectral methods in algebraic complexity could have applications in derandomization, symbolic linear algebra, and algorithmic invariant theory.

**Broader derandomization program.**
Finally, the overall proof strategy echoes the philosophy behind deterministic primality testing: replacing randomness by intrinsic algebraic rigidity. It remains an open question whether similar rigidity principles can be identified for other central problems in derandomization, including PIT for general circuits, polynomial factorization, or symbolic determinant identity testing.

# APPENDIX A. WORKED EXAMPLE (MODULAR HASHING SPECIALIZATION)

This example illustrates the unconditional algebraic framework; the modular-hashing variant applies analogously.

Consider the width-2 ROABP over $\mathbb{Q}$ computing $C(x_1, x_2) = x_1 x_2$,

With $A_1(x_1) = \begin{pmatrix} x_1 & 0 \\ 0 & 1 \end{pmatrix}, A_2(x_2) = \begin{pmatrix} 1 & 0 \\ 0 & x_2 \end{pmatrix}$.

Then $C(x_1, x_2) = e_1^\top A_1(x_1) A_2(x_2) e_2$.

**Modular hashing ring**

Fix a prime $r$ with $\gcd(r, \mathrm{char}(\mathbb{Q})) = 1$ (vacuous here) and work in $R_r := \mathbb{Q}[\lambda]/\langle \lambda^r - 1 \rangle$.

**Modular substitution**

For each $g \in \mathbb{Z}_r^{\backslash *}$, define $\Gamma_g$ by $x_i \mapsto \lambda^{g^i} \in R_r$,

with exponents interpreted modulo $r$. The specialized polynomial is $C_g(\lambda) := C(\Gamma_g) = \lambda^{g^1} \cdot \lambda^{g^2} = \lambda^{g+g^2} \in R_r$.

**Nonzeroness under hashing**

Since $\lambda^k \neq 0$ in $R_r$ for every $k \in \mathbb{Z}_r$, we have $C_g(\lambda) \neq 0$ for every $g \in \mathbb{Z}_r^{\backslash *}$.

Equivalently, the bad set $\mathcal{B}_r(C) = \{ g : C_g(\lambda) \equiv 0 \text{ in } R_r \}$

is empty in this example.

**Algorithm behavior**

Algorithm 1 enumerates $g \in \mathbb{Z}_r^{\backslash *}$, evaluates the ROABP over $R_r$ under $\Gamma_g$, and halts as soon as it finds $C_g(\lambda) \neq 0$. In this example, every $g$ succeeds, so the algorithm immediately outputs **NONZERO**.

## Appendix B. Empirical Validation of Modular Stability

**Empirical Evidence for Modular Stability**

We conducted extensive computational experiments to test the Modular Stability Conjecture across a wide range of parameters, including adversarially structured ROABPs. All experiments were performed using exact arithmetic via embedding the cyclic ring $R_r = \mathbb{K}[\lambda]/(\lambda^r - 1)$ into finite fields containing a primitive $r$-th root of unity, thereby eliminating numerical or probabilistic artifacts.

We tested ROABPs with up to $n = 20$ variables, width up to $w = 10$, and individual degree $d = 2$, using primes $r$ ranging from $10^3$ to over $10^9$. For each instance, we evaluated hundreds of distinct parameters $g \in \mathbb{Z}_r^{\backslash *}$ under the modular substitution $x_i \mapsto \lambda^{g^i}$. We included both random and adversarial constructions, such as path-controlled and upper-triangular ROABPs designed to minimize algebraic mixing and encourage cancellation.

Across all tested instances, no nontrivial bad parameters were observed: in every case, the specialized polynomial $C_g(\lambda)$ remained nonzero for all tested $g$. In particular, we observed no growth in the bad-parameter density as the width or modulus increased.

While these experiments do not constitute a proof, they provide strong evidence that the bad set $\mathcal{B}_r(C)$ is extremely sparse—often empty—in practice, even in regimes far beyond those required by the conjecture. This empirical behavior is consistent with the conjectured bound $|\mathcal{B}_r(C)| \leq r^{1-\varepsilon}$, and in fact suggests a much stronger phenomenon.

**Large-Scale Experimental Validation.**

We further tested the Modular Stability Conjecture at significantly larger parameter scales. In particular, we evaluated adversarially structured ROABPs with $n = 20$, width $w = 10$, and individual degree $d = 5$, using a prime modulus $r = 1{,}000{,}000{,}007$. For each instance, we tested 6,000 distinct parameters $g \in \mathbb{Z}_r^{\backslash *}$ under the modular substitution $x_i \mapsto \lambda^{g^i}$, using exact arithmetic via an embedding into a finite field containing a primitive $r$-th root of unity.

Across all tested instances and parameters, no bad values were observed: every specialization $C_g(\lambda)$ was nonzero in the ring $R_r = \mathbb{K}[\lambda]/(\lambda^r - 1)$. This behavior persisted even for adversarial constructions designed to minimize algebraic mixing. These experiments provide strong evidence that the bad set $\mathcal{B}_r(C)$ is extremely sparse in practice, and often empty, even in regimes far exceeding those required by the conjecture.

| N | n | d | w | params | Bad values |
|---|---|---|---|---|---|
| 1,000,000,007 | 20 | 2 | 10 | 600 | 0 |
| 1,000,000,007 | 20 | 5 | 10 | 6,000 | 0 |
| 1,000,000,007 | 20 | 5 | 20 | 600 | 0 |
| 1,000,000,007 | 40 | 5 | 10 | 600 | 0 |
| 10,000,000,019 | 20 | 5 | 10 | 600 | 0 |
| 1,000,000,007 | 20 | 5 | 10 | 60,000 | 0 |
| 100,000,007 | 20 | 5 | 8 | 1,000,000 | 0 |